\documentclass[a4paper,11pt, table]{article}
\pdfoutput=1

\usepackage{jheppub}

\usepackage[T1]{fontenc} 
\usepackage{epsf,amsmath,amssymb,graphicx,dcolumn}
\usepackage{scalefnt}
\usepackage{hyperref}
\usepackage{cleveref,etoolbox,color,soul}
\usepackage{listings,booktabs}
\usepackage[utf8]{inputenc}
\usepackage{epsfig}
\usepackage{graphicx}
\usepackage[colorinlistoftodos]{todonotes}
\usepackage{multirow}
\usepackage{bbold}
\usepackage{bm}
\usepackage{amssymb}
\usepackage{amsmath}
\usepackage{subcaption}

\newcommand{\Ml}{\mathbf{M}_{\ell}}
\newcommand{\M}{\mathbf{M}}
\newcommand{\U}{\mathbf{U}}
\newcommand{\V}{\mathbf{V}}

\newcommand{\Y}{\mathbf{Y}}

\def\EW{$\mathrm{SU(2)}_L \otimes \mathrm{U(1)}_Y$}

\def\MR{\mathbf{M}_R}
\def\Yl{\mathbf{Y}_\ell}
\def\Y{\mathbf{Y}}

\def\YD{\mathbf{Y}_D}
\def\YR{\mathbf{Y}_R}
\def\YRp{\mathbf{Y}_R'}

\def\Mnu{\mathbf{M}_\nu}
\def\MD{\mathbf{M}_D}
\def\Mnuh{\widehat{\mathbf{M}}_\nu}
\def\Ml{\mathbf{M}_\ell}
\def\Ur{\mathbf{U}_R}

\def\U{\mathbf{U}}

\newcolumntype{K}[1]{>{\centering\arraybackslash}m{#1}}
\def\gsim{\raise0.3ex\hbox{$\;>$\kern-0.75em\raise-1.1ex\hbox{$\sim\;$}}}
\def\lsim{\raise0.3ex\hbox{$\;<$\kern-0.75em\raise-1.1ex\hbox{$\sim\;$}}}

\definecolor{linkcolor}{rgb}{0,0,0.5}

\def\Y{\mathbf{Y}}

\title{\boldmath Scalar-singlet assisted leptogenesis with CP violation from the vacuum}

\author[a]{D. M. Barreiros,}
\author[a]{H. B. Câmara,}
\author[b,a]{R. G. Felipe,}
\author[a]{and F. R. Joaquim}

\affiliation[a]{Departamento de F\'{\i}sica and CFTP, Instituto Superior T\'ecnico, Universidade de Lisboa, Lisboa, Portugal}
\affiliation[b]{ISEL - Instituto Superior de Engenharia de Lisboa, Instituto Polit\'ecnico de Lisboa, Rua Conselheiro Em\'{\i}dio Navarro, 1959-007 Lisboa, Portugal}

\emailAdd{debora.barreiros@tecnico.ulisboa.pt}
\emailAdd{henrique.b.camara@tecnico.ulisboa.pt}
\emailAdd{ricardo.felipe@isel.pt}
\emailAdd{filipe.joaquim@tecnico.ulisboa.pt}

\abstract{
In the vanilla type-I seesaw leptogenesis scenario, CP violation required to generate the lepton asymmetries in the heavy Majorana neutrino decays stem from complex Dirac-type Yukawa couplings. In this paper we explore the case in which that CP violation originates from the vacuum expectation value of a complex scalar singlet at a very high scale. This non-trivial CP-violating phase can be successfully communicated to the low-energy neutrino sector via the heavy neutrino portal. The new scalar-singlet degrees of freedom generate new contributions to the CP asymmetries relevant for leptogenesis not only at the one-loop level but also through tree-level three-body decays. These are computed here for an arbitrary number of heavy neutrinos, Higgs doublets and scalar singlets. We also take into account the new decays and scattering processes that enter the unflavoured Boltzmann equations governing the heavy-neutrino particle densities and the $(B-L)$-asymmetry evolution. Having established the framework of interest, we present a simple model with two RH neutrinos, two Higgs doublets and a complex scalar singlet, supplemented with a $\mathcal{Z}_8$ flavour symmetry. This symmetry minimises the number of free parameters without compromising the possibility of spontaneous CP violation and compatibility with neutrino data. In fact, the only viable $\mathcal{Z}_8$ charge assignment shows a preference for a non-trivial spontaneous CP-violating phase, which in turn leads to a non-vanishing CP asymmetry due to the direct link between high- and low-energy CP violation. An interesting feature of this simple setup is that the usual wave and vertex type-I seesaw contributions to the CP asymmetry vanish due to the $\mathcal{Z}_8$ symmetry. Thus, the observed baryon-to-photon ratio can be explained thanks to the new couplings among the heavy neutrinos and the new scalar degrees of freedom.}

\begin{document}
\maketitle
\flushbottom

\section{Introduction}
\label{sec:intro}

The observation of neutrino oscillations~\cite{McDonald:2016ixn,Kajita:2016cak} requires the existence of neutrino masses and lepton mixing, thus providing evidence for physics beyond the Standard Model~(SM). The ever growing neutrino experimental programme has been shedding light on neutrino properties with neutrino oscillation experiments measuring with improving precision the neutrino mass-squared differences, mixing angles and the Dirac charge-parity (CP) violating phase $\delta$. Global fits of the data provide up-to-date values of neutrino observables, being some of them known with subpercent precision~\cite{deSalas:2020pgw,Esteban:2020cvm,Capozzi:2021fjo}. However, several neutrino-related questions remain unanswered. For instance, we do not know whether neutrinos are Majorana or Dirac particles, nor their mass ordering or absolute mass scale. Moreover, there is no solid confirmation that CP is violated in the lepton sector. Currently, while some tension between NO$\nu$A and T2K results exist regarding the Dirac CP phase for a normal neutrino-mass ordering, maximal CP violation seems to be preferred for an inverted spectrum (for a recent review see ref.~\cite{Rahaman:2022rfp}). In the next decades, long-baseline neutrino experiments like DUNE~\cite{DUNE:2016hlj} and Hyper-Kamiokande~\cite{Hyper-Kamiokande:2018ofw} will provide much more information on leptonic CP violation~(LCPV)~\cite{Branco:2011zb}. At the same time, numerous searches for neutrinoless double beta decay, sensitive to Majorana CP violation, will be crucial to probe on the particle nature of neutrinos (see refs.~\cite{Bilenky:2014uka,DellOro:2016tmg,Dolinski:2019nrj} for recent reviews on this subject).

Explaining (and testing) the origin of neutrino masses and mixing remains a challenging problem. From the theory viewpoint, the fact that neutrinos are at least six orders of magnitude lighter than the electron can be successfully accommodated within the framework of the seesaw mechanism~\cite{Minkowski:1977sc,Gell-Mann:1979vob,Yanagida:1979as,Glashow:1979nm,Mohapatra:1979ia,Konetschny:1977bn,Cheng:1980qt,Lazarides:1980nt,Schechter:1980gr,Mohapatra:1980yp,Magg:1980ut}. The type-I seesaw stands out among its various realisations which, in its minimal version, requires two right-handed~(RH) neutrinos~\cite{Frampton:2002qc,Ibarra:2003up,Harigaya:2012bw,Rink:2016vvl,Shimizu:2017fgu,Barreiros:2018ndn,Barreiros:2018bju,Barreiros:2020mnr}. The heaviness of these new states leads to neutrino mass suppression and their Yukawa couplings with SM lepton and the Higgs doublets provide potential sources of LCPV. In general, the number of parameters in the full Lagrangian of the SM extended with RH neutrinos exceeds the number of low-energy neutrino mass and mixing observables. Hence, in order to obtain testable low-energy predictions for lepton mixing and LCPV one can consider, for instance, theoretical frameworks with maximally-restrictive neutrino mass matrices combined with Abelian flavour symmetries~\cite{Grimus:2004hf,Dighe:2009xj,Adhikary:2009kz,Dev:2011jc,GonzalezFelipe:2014zjk,Cebola:2015dwa,Samanta:2015oqa,Kobayashi:2018zpq,Rahat:2018sgs,Nath:2018xih,Correia:2019vbn,Camara:2020efq}.

The SM also fails at explaining the observed baryon asymmetry of the Universe~(BAU). Indirect astrophysical observations, namely, anisotropies in the cosmic microwave background, large scale formation data and Big Bang nucleosynthesis, among others, indicate that there is more matter than antimatter in the Universe. The most recent data from the Planck satellite sets the current value for the baryon-to-photon ratio at $\eta_B^0 = (6.12 \pm 0.04) \times 10^{-10}$  at $68\%$ confidence level (CL)~\cite{Planck:2018vyg}. To generate a baryon asymmetry from a matter state initially symmetric, one needs to fulfill the three Sakharov conditions: i) C and CP violation, ii) B violation and iii) departure from thermal equilibrium~\cite{Sakharov:1967dj}. It turns out that the amount of CP violation in the SM is too small to successfully generate the observed BAU via electroweak baryogenesis~\cite{Gavela:1993ts,Gavela:1994ds,Gavela:1994dt}. This motivates the study of SM extensions with new sources of explicit or spontaneous CP violation~(SCPV), and new mechanisms to generate the BAU. In fact, within the seesaw paradigm, the excess of matter over antimatter can be explained through the leptogenesis mechanism~\cite{Fukugita:1986hr} (for reviews on leptogenesis see, e.g., refs.~\cite{Buchmuller:2004nz,Davidson:2008bu,Fong:2012buy,Hambye:2012fh} and for detailed analyses in the context of the minimal type-I seesaw, see refs.~\cite{Frampton:2002qc,Ibarra:2003up,GonzalezFelipe:2003fi,Joaquim:2005zv,Branco:2005jr,Abada:2006ea,Harigaya:2012bw,Zhang:2015tea,Siyeon:2016wro,Rink:2016vvl,Geib:2017bsw,Achelashvili:2017nqp,Shimizu:2017fgu,Shimizu:2017vwi,Covi:1996wh,Antusch:2011nz,Barreiros:2018ndn,Barreiros:2020mnr}). This is realised via the out-of-equilibrium lepton-number violating~(LNV) decays of heavy neutrinos in the early Universe, generating a lepton asymmetry which is subsequently converted into a baryon asymmetry by non-perturbative sphalerons~\cite{Kuzmin:1985mm}. 

In contrast to SM fermions, bare RH Majorana neutrino mass terms are invariant under the SM gauge group. Still, one can envisage scenarios where heavy neutrino masses are generated dynamically by adding scalar fields coupled to the RH neutrinos. After acquiring a non-zero vacuum expectation value~(VEV), heavy masses could be generated by those VEVs as, e.g., in Majoron models~\cite{Chikashige:1980qk,Chikashige:1980ui,Gelmini:1980re}. The existence of new scalar degrees of freedom coupled to heavy Majorana neutrinos induce new contributions to the CP asymmetries relevant for the generation of a lepton asymmetry in leptogenesis scenarios~\cite{Pilaftsis:2008qt,AristizabalSierra:2014uzi,LeDall:2014too,Alanne:2017sip,Alanne:2018brf}.

In this work we consider the case in which leptogenesis is assisted by complex scalar singlets. The latter may acquire complex VEVs which, being the sole source of CP violation, provide a common origin for CP-violating effects at low and high energies~\cite{Branco:2003rt}. We investigate the possibility that the spontaneous breaking of CP occurs at a scale above the leptogenesis scale, in such a way that the complex singlet VEVs give rise to complex RH neutrino mass terms. In the fermion and scalar mass eigenstate basis, this leads to non-trivial CP-violating scalar-heavy neutrino interactions which, in turn, trigger new contributions to the CP asymmetries in heavy neutrino decays. On the other hand, the evolution of particle number densities controlled by Boltzmann equations~(BEs) is also affected by the presence of those new interactions. The whole setup is illustrated with a concrete model based on a SM extension with 2RH neutrinos, two Higgs doublets and a complex scalar singlet, supplemented with a discrete $\mathcal{Z}_8$ (flavour) symmetry. 

The paper is organised as follows. In section~\ref{sec:framework}, we review the general framework of the type-I seesaw mechanism, starting with an arbitrary number of Higgs doublets $n_H$ and complex scalar singlets $n_S$. It is shown how CP violation generated from the singlet VEVs can be successfully communicated to the lepton sector. The implementation of thermal leptogenesis in this scenario is then analysed in section~\ref{sec:leptogenesis}. We present the new CP-asymmetry contributions arising from the presence of scalar singlets and study the unflavoured BEs taking into account the relevant decays and scattering processes. Having established the theoretical framework, in section~\ref{sec:model} we illustrate our idea in a minimal model, showing that SCPV can be simultaneously responsible for LCPV and successfully generate the observed value of the BAU. Finally, our concluding remarks are presented in section~\ref{sec:concl}. Generalities and notation regarding BEs, as well as the analytical expressions for the scattering cross sections, can be found in the appendices.

\section{Type-I seesaw and high-energy SCPV}
\label{sec:framework}

In the class of type-I seesaw models we are interested in, there are essentially two ways of breaking CP: i) explicitly by considering gauge-invariant complex Yukawa couplings and/or bare mass terms and ii) spontaneously through the complex VEVs of some scalar fields. In the latter case, the spin-0 complex fields can be singlets, doublets, triplets, or even more complicated multiplets, if the underlying symmetry groups are larger than the SM one. In this work, we will consider SM extensions with $n_R$ RH neutrinos $\nu_R$, $n_H$ scalar doublets $\Phi_a$ ($a=1, \cdots, n_H$) and $n_S$ complex scalar singlets $S_k$ ($k=1, \cdots, n_S$). Within this general setup, the Yukawa and mass terms allowed by the gauge symmetry are
\begin{align}
- \mathcal{L}_{\text{Yuk.}} =\overline{\ell}_L{\Yl^a} \Phi_a e_R+\overline{\ell}_L \YD^{a \ast} \tilde{\Phi}_a\nu_R+\dfrac{1}{2}\overline{\nu_R}\left(\MR^0 + \YR^k S_k+ \YRp^k S^*_k \right)\nu_R^c+\text{H.c.} \; ,
\label{eq:LYuk}
\end{align}
where $\ell_L=(\nu_L\; e_L)^T$ and $e_R$ denote the SM left-handed doublet and RH singlet charged-lepton fields, respectively; $\nu_R = ( \nu_{R\, 1}, \cdots, \nu_{R\, n_R})^T$ and $\nu_R^c \equiv C\, \overline{\nu_R}^T $, being $C$ the charge conjugation matrix. $\Yl^a$, $\YD^{a \ast}$, $\YR^k$ and $\YRp^k$ are, respectively,  $3\times 3$, $3\times n_R$ and $n_R \times n_R$ complex Yukawa matrices, being the latter two symmetric. Bare RH neutrino masses are denoted by the $n_R \times n_R$ symmetric matrix $\MR^0$. The doublet and singlet  scalar fields $\Phi_a$ and $S_k$ are defined in the usual form:
\begin{align}
\Phi_a =\begin{pmatrix}
\phi^{+}_a \\
\phi^0_a
\end{pmatrix}= \frac{1}{\sqrt{2}}  \begin{pmatrix}
 \sqrt{2} \phi^{+}_a \\
 v_a e^{i \varphi_a} + \phi^0_{\text{R} a} + i \phi^0_{\text{I} a}
\end{pmatrix} \; , \;  S_k = \frac{1}{\sqrt{2}}\left( u_k\, e^{i \theta_k} + S_{\text{R} k} + i S_{\text{I} k}\right) \, ,
\label{eq:scalardef}
\end{align}
with $\tilde{\Phi}_a=i\tau_2 \Phi^\ast_a$; $\tau_2$ is the Pauli matrix and the doublet VEVs are normalised as $v^2 = \sum_a |v_a|^2$ with $v\simeq 246$ GeV. The $S_k\,\overline{\nu_R}\,\nu_R^c$ Yukawa interactions give rise to mass terms for the $\nu_R$'s if the scalar singlets develop non-zero VEVs $u_k\neq0$. After electroweak symmetry breaking~(EWSB), the full fermion-mass Lagrangian is
\begin{equation}
\begin{split}
-\mathcal{L}_{\text{mass}} & = \overline{e_{L}}\, \Ml\, e_{R} + \overline{\nu_L}\, \MD^\ast\, \nu_R + \dfrac{1}{2}\overline{\nu_R}\, \MR\,  \nu_R^c + \text{H.c.} \; ,
\end{split}
\label{eq:Lmass}
\end{equation}
where $\mathbf{M}_{\ell}$, $\mathbf{M}_{D}$ and $\mathbf{M}_{R}$ are the charged-lepton, Dirac neutrino and RH neutrino mass matrices, respectively given by:
\begin{align}
   \Ml = \frac{v_a}{\sqrt{2}} e^{i \varphi_a} \Yl^a \; , \; \mathbf{M}_{D}^\ast = \frac{v_a}{\sqrt{2}} e^{- i \varphi_a} \YD^{a \ast} \; , \; \MR = \MR^0 + \frac{u_k}{\sqrt{2}} \left(\YR^k e^{i \theta_k} + \YRp^k e^{- i \theta_k}\right) \, ,
\label{eq:massyuk}
\end{align}
where sums over repeated indices are implicit. Defining the $n_f = 3+n_R$ component vector $N_{L}=\left(\nu_{L}, \nu^c_{R} \right)^{T}$ in flavour space, we can write $\mathcal{L}_{\text{mass}} $ as
\begin{align}
-\mathcal{L}_{\text{mass}} =\overline{e_{L}}\, \Ml\, e_{R}+\frac{1}{2} \overline{N_{L}^c} \bm{\mathcal{M}} N_{L} + \text{H.c.} \; , \; 
\bm{\mathcal{M}} = 
\begin{pmatrix} 0 & \MD \\
\MD^{T} & \MR
\end{pmatrix} \, .
\label{eq:bigm}
\end{align}

The charged-lepton mass matrix is bidiagonalised through the unitary transformations $e_{L}  \rightarrow \mathbf{V}_{L}\, e_{L}, \ e_{R}  \rightarrow \mathbf{V}_{R}\, e_{R}$, so that
\begin{equation}
\begin{aligned}
 \mathbf{V}_{L}^{\dagger} \mathbf{M}_{\ell} \mathbf{V}_{R} = \mathbf{d}_{\ell} = \text{diag}\left(m_e, m_{\mu}, m_{\tau}\right),
\end{aligned}
\label{diagcharg}
\end{equation}
with $m_{e,\mu,\tau}$ denoting the physical charged-lepton masses. For a given $\Ml$, the unitary matrices $\mathbf{V}_{L}$ and $\mathbf{V}_{R}$ are determined through the standard procedure, i.e. by diagonalising the Hermitian matrices $\mathbf{H}_{\ell} = \mathbf{M}_{\ell} \mathbf{M}_{\ell}^{\dagger}$ and $\mathbf{H}_{\ell}^\prime = \mathbf{M}_{\ell}^{\dagger} \mathbf{M}_{\ell}$ as
\begin{equation}
\mathbf{V}^{\dagger}_{L} \mathbf{H}_{\ell} \mathbf{V}_{L} = \mathbf{d}^2_{\ell} = \text{diag}\left(m_e^2 , m_{\mu}^2, m_{\tau}^2\right) \; , \; \mathbf{V}^{\dagger}_{R} \mathbf{H}_{\ell}^\prime \mathbf{V}_{R} = \mathbf{d}^2_{\ell} = \text{diag}\left(m_e^2 , m_{\mu}^2, m_{\tau}^2\right).
\label{eq:VLmix}
\end{equation}

In the seesaw approximation limit, when $\mathbf{M}_D \ll \mathbf{M}_{R}$, the neutrino mass matrix $\bm{\mathcal{M}}$ of eq.~\eqref{eq:bigm} can be block-diagonalised yielding the well-known $3\times 3$ effective light neutrino mass matrix,
\begin{equation}
\Mnu = - \MD \MR^{-1} \MD^T \,.
\label{eq:typeI}
\end{equation}
Hence, the active neutrinos acquire mass via the type-I seesaw mechanism. The above matrix can be diagonalised through a unitary rotation $\nu_{L}  \rightarrow \mathbf{U}_{\nu}\, \nu_{L}$, satisfying
\begin{equation}
\begin{aligned}
\mathbf{U}_{\nu}^{ T}\, \Mnu\, \mathbf{U}_{\nu}  = \mathbf{d}_{\nu} = \text{diag}\left(m_1, m_2, m_3\right),
\end{aligned}
\label{eq:Unudef}
\end{equation}
where $m_{1,2,3}$ are the real and positive light neutrino masses. The unitary matrix $\mathbf{U}_{\nu}$ is obtained by diagonalising the Hermitian matrix $\mathbf{H}_{\nu} = \Mnu \Mnu^{\dagger}\,$,
\begin{equation}
\mathbf{U}^{\dagger}_{\nu} \mathbf{H}_{\nu} \mathbf{U}_{\nu}  = \mathbf{d}^2_{\nu} = \text{diag}\left(m_1^2 , m_{2}^2, m_{3}^2\right).
\end{equation}
As a result, the unitary lepton mixing matrix is given by
\begin{equation}
   \mathbf{U} = \mathbf{V}_{L}^{\dagger} \mathbf{U}_{\nu}\,,
   \label{eq:leptonmix}
\end{equation}
after performing the rotation to the charged-lepton mass basis. Finally, the mass matrix~$\MR$ can be diagonalised through a unitary rotation of the heavy neutrino fields, satisfying
\begin{equation}
\begin{aligned}
{\nu_{R}}_{i} \rightarrow (\Ur)_{i j} \, N_j\;,\;\Ur^{\dagger}\, \MR\, \Ur^\ast = \mathbf{d}_{M} = \text{diag}\left(M_1, \cdots, M_{n_R}\right),
\end{aligned}
\end{equation}
yielding $n_R$ heavy neutrinos $N_j$ with real and positive masses $M_j$. Defining the Hermitian matrix $\mathbf{H}_{R} = \MR \MR^{\dagger}\,$, the unitary matrix $\Ur$ is obtained from 
\begin{equation}
\Ur^{\dagger} \mathbf{H}_{R} \Ur = \mathbf{d}^2_{M} = \text{diag}\left(M_1^2, \cdots, M_{n_R}^2\right).
\label{eq:heavymix}
\end{equation}

As for the scalar potential, the new trilinear and quartic terms are
\begin{align}
V(\Phi,S) & \supset \mu_{a b, i} (\Phi_a^\dagger \Phi_b) S_i + \lambda_{a b, i k} (\Phi_a^\dagger \Phi_b) S_i S_k^\ast + \lambda_{a b, i k}^\prime (\Phi_a^\dagger \Phi_b) S_i S_k \nonumber \\  
&+ \mu_{i j k} S_i^\ast S_j S_k + \mu^\prime_{i j k} S_i S_j S_k \nonumber \\ 
& + \lambda_{i j k l} S_i^\ast S_j^\ast S_k S_l + \lambda_{i j k l}^\prime S_i^\ast S_j S_k S_l + \lambda_{i j k l}^{\prime \prime} S_i S_j S_k S_l + \text{H.c.}\; ,\label{eq:cubicscalar} \end{align} 
where the mass (dimensionless) $\mu$ ($\lambda$) are complex parameters. The scalar interactions stemming from \eqref{eq:cubicscalar} are of special interest since they will induce new contributions to the CP asymmetries in $N_i$ decays, and to the scattering processes entering the BEs (see section~\ref{sec:leptogenesis}).\footnote{More details on the full scalar potential will be given for the specific case of two Higgs doublets and one scalar singlet in section~\ref{sec:model}. For simplicity, hereafter we  neglect the quartic terms since they will not play a relevant role in our analysis.}

In this work, we consider the scenario in which the singlets acquire complex VEVs at energies well above the EW scale $v$. At temperatures $T \gg v$, the Higgs doublets are VEVless, i.e. the EW symmetry is still unbroken. By imposing CP conservation at the Lagrangian level, all couplings in the Yukawa and scalar sectors are real. Consequently, under certain conditions, CP may be broken if the singlet fields $S_k$ develop complex VEVs, meaning that the sole source of CP violation in our framework comes from SCPV occuring at very high energies. This CP violation can be transmitted in a nontrivial way to the neutrino sector through the heavy neutrino-scalar portal $\overline{\nu_R}\, \nu_R^c\, S$, provided that the VEV phases appearing in the mass matrix $\M_R$, defined in eq.~\eqref{eq:massyuk}, are not removable by field redefinitions. The link with low-energy LCPV effects is established when the Higgs doublets acquire non-zero VEVs and the EW symmetry is spontaneously broken. At this stage, the charged-lepton $\Ml$ and Dirac-type neutrino $\MD$ mass matrices are generated, giving masses to the SM leptons and to the light neutrinos via the type-I seesaw mechanism. Since the complex matrix $\MR$ enters the expression for the effective light neutrino mass matrix $\mathbf{M}_{\nu}$ in eq.~\eqref{eq:typeI}, non-trivial Dirac and Majorana phases may appear in the lepton mixing matrix. As we will show in the next section, besides explaining LCPV, vacuum CP violation can also lead to non-vanishing CP asymmetries in leptogenesis.

At the leptogenesis scale, the scalar degrees of freedom contained in the singlets will be massive with a typical mass of order $u \gg v$. In fact, there will be $2 n_S$ scalar mass eigenstates $h_i$ with mass matrix $\bm{\mathcal{M}}_S$, being the mixing with the weak states given by
\begin{equation}
(S_{\text{R} 1}, \cdots, S_{\text{R} n_S}, S_{\text{I} 1}, \cdots, S_{\text{I} n_S})^T= \mathbf{V} (h_1, \cdots, h_{2 n_S})^T\;,
\label{eq:mixS}
\end{equation}
where $\mathbf{V}$ is a $2 n_S \times 2 n_S$ orthogonal matrix, such that
\begin{align}
\mathbf{V}^T \bm{\mathcal{M}}_S^2 \mathbf{V} = \mathbf{d}^2_S = \text{diag}(m_{h_1}^2, \cdots, m_{h_{2 n_S}}^2) \; .
\label{eq:mixingscalar}
\end{align}
As usually done in leptogenesis calculations, we will consider that, before EWSB, the heavy Majorana neutrinos $N_i$ and scalars $h_k$ are much heavier than the scalars stemming from the Higgs doublets $\Phi_a$, i.e. $m_{\Phi_a}\ll M_i, m_{h_k}$. For this reason, we neglect $m_{\Phi_a}$ in our calculations. Note also that the SM fermions are massless in the symmetric phase.

For computational purposes, we define some of the Yukawa and scalar couplings in the flavour-diagonal basis of charged-lepton Yukawa couplings, and on the mass basis of heavy neutrinos and scalar fields. Namely, for the $\overline{\ell_L} N \Phi_a$, $N N h_k$, $(\Phi_a^\dagger \Phi_b) h_k$ and $h_i h_j h_k$ couplings we now have\footnote{Notice that here we are already putting the quartic couplings of eq.~\eqref{eq:cubicscalar} to zero. }
\begin{align}
\mathbf{Y}^{a\ast}\overline{\ell_L} N \Phi_a: \mathbf{Y}^{a\ast} & = \mathbf{V}_L^\dagger \YD^{a \ast} \Ur \; , \; \mathbf{H}^a = \mathbf{Y}^{a \dagger} \mathbf{Y}^{a} \; , \label{eq:YHdef} \\
 \mathbf{\Delta}^k NNh_k: \mathbf{\Delta}^k &= \frac{1}{2 \sqrt{2}} \sum_{j=1}^{n_S} \Ur^\dagger \left[ \mathbf{V}_{j k} (\YR^j + \YRp^j) + i \mathbf{V}_{j+n_S k} (\YR^j - \YRp^j) \right] \Ur^\ast , \label{eq:Deltadef} \\
 \tilde{\mu}_{a b, k}(\Phi_a^\dagger \Phi_b) h_k: \tilde{\mu}_{a b, k} &= \sqrt{2} \sum_{j=1}^{n_S} \left[\text{Re}\left(\mu_{a b, j}\right) \mathbf{V}_{j k} - \text{Im}\left(\mu_{a b, j}\right) \mathbf{V}_{j+n_S k}\right]\,,
\label{eq:muCPdef}
\\
 \tilde{\mu}_{i j k} h_i h_j h_k: \tilde{\mu}_{i j k} &= \frac{1}{\sqrt{2}} \sum_{l,p,q=1} \big[  \text{Re}\left(\mu_{lpq} + \mu_{lpq}^\prime \right)  \mathbf{V}_{l i} \left( \mathbf{V}_{p j} \mathbf{V}_{q k} - \mathbf{V}_{p+n_S j} \mathbf{V}_{q+n_S k} \right)\nonumber \\ 
 & + \text{Re}\left(\mu_{lpq} - \mu_{lpq}^\prime \right) \mathbf{V}_{l+n_S i} \left( \mathbf{V}_{p+n_S j} \mathbf{V}_{q k} + \mathbf{V}_{p j} \mathbf{V}_{q+n_S k} \right) \nonumber \\ 
 & - \text{Im}\left(\mu_{lpq} + \mu_{lpq}^\prime \right) \mathbf{V}_{l i} \left( \mathbf{V}_{p+n_S j} \mathbf{V}_{q k} + \mathbf{V}_{p j} \mathbf{V}_{q+n_S k} \right) \nonumber \\ 
 & + \text{Im}\left(\mu_{lpq} - \mu_{lpq}^\prime \right) \mathbf{V}_{l+n_S i} \left( \mathbf{V}_{p j} \mathbf{V}_{q k} - \mathbf{V}_{p+n_S j} \mathbf{V}_{q+n_S k} \right) \big]\,,
\label{eq:muhkCPdef}
\end{align}
where the field rotations were performed in eqs.~\eqref{eq:LYuk} and \eqref{eq:cubicscalar} using the unitary matrices $\mathbf{V}_L$, $\Ur$ and $\mathbf{V}$ given in eqs.~\eqref{eq:VLmix}, \eqref{eq:heavymix} and \eqref{eq:mixingscalar}, respectively.

\section{Leptogenesis assisted by complex scalar singlet}
\label{sec:leptogenesis}

The BAU is quantified by the baryon-to-photon ratio
\begin{align}
    \eta_B \equiv \frac{n_B - n_{\bar{B}}}{n_\gamma} \; ,
\end{align}
with $n_B$, $n_{\bar{B}}$ and $n_\gamma$ being, respectively, the number densities of baryons, antibaryons and photons. The present value for $\eta_B$, obtained from the combined Planck
TT,TE,EE+ lowE+lensing data is~\cite{Planck:2018vyg},
\begin{align}
    \eta_B^0 = (6.12 \pm 0.04) \times 10^{-10} \; ,
    \label{eq:etab0}
\end{align}
at 68\% CL. In the type-I seesaw framework, leptogenesis proceeds via the out-of-equilibrium decays of the heavy neutrinos $N_i$ in the early Universe. The resulting lepton asymmetry is then partially converted into a baryon asymmetry through the $(B+L)$-violating sphaleron transitions, leading to~\cite{Antusch:2011nz}
\begin{align}
    \eta_B = a_{\text{sph}} \frac{N_{B-L}^f}{N_{\gamma}^{\text{rec}}} \simeq 9.40 \times 10^{-3} N_{B-L}^f \; ,
    \label{eq:etaB}
\end{align}
where, for $3$ fermion generations, $a_{\text{sph}} = (24 + 4 n_H)/(66 + 13 n_H)$ is the sphaleron conversion factor~\cite{Khlebnikov:1988sr,Harvey:1990qw}, $N_{B-L}^f$ is the final asymmetry calculated in a comoving volume and $N_{\gamma}^{\text{rec}} \simeq 37.01$ is the number of photons in the same comoving volume at the recombination temperature. The approximation in \eqref{eq:etaB} corresponds to $n_H = 2$ which will be the case under study in section~\ref{sec:model}.

\begin{figure}[t!]
    \centering
     \begin{subfigure}[b]{0.4\textwidth}
         \centering
         \includegraphics[scale=0.9,trim={5cm 23.5cm 12cm 1.0cm},clip]{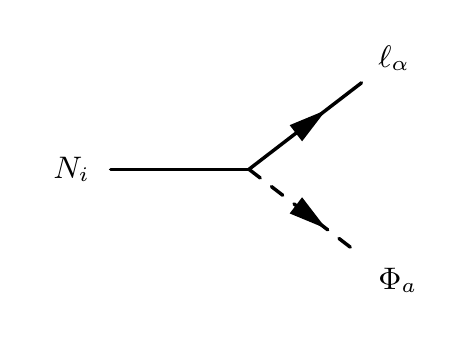}
         \caption{}
         \label{fig:treetypeI}
     \end{subfigure}
     \hspace{1cm}
     \begin{subfigure}[b]{0.4\textwidth}
         \centering
         \includegraphics[scale=0.9,trim={11.0cm 23.5cm 2cm 1.0cm},clip]{diagrams_CPasymmetry.pdf}
         \caption{}
         \label{fig:vertextypeI}
     \end{subfigure}\\
     \begin{subfigure}[b]{0.9\textwidth}
         \centering
         \includegraphics[scale=0.9,trim={4.0cm 20.3cm 3.0cm 4.0cm},clip]{diagrams_CPasymmetry.pdf}
         \caption{}
         \label{fig:wavetypeI}
     \end{subfigure} \\
    \caption{Usual diagrams contributing to the CP asymmetry in $N_i \rightarrow \ell_\alpha \Phi_a$ through interference. (a) Tree-level contribution $ \sim \mathcal{O}(Y^2)$. (b) Vertex contribution $\sim \mathcal{O}(Y^4)$. (c) Wavefunction contributions $\sim \mathcal{O}(Y^4)$.}
    \label{fig:usualcontribution_CPasym}
\end{figure}

A key ingredient in the generation of the BAU within the (standard) leptogenesis framework are the CP asymmetries $\varepsilon_{i \alpha}^a$ produced in the heavy neutrino decays $N_i \rightarrow \Phi_a \ell_\alpha$, where $\ell_\alpha$ represents one lepton of flavour $\alpha= e, \mu, \tau$ and $\Phi_a$ denotes a scalar-field component of the doublet with $a=1,...,n_H$. The CP asymmetries $\varepsilon_{i \alpha}^a$ are defined as follows~\cite{Covi:1996wh}
\begin{align}
    \varepsilon_{i \alpha}^a = \dfrac{\Gamma(N_i \rightarrow \Phi_a \ell_\alpha) - \Gamma(N_i \rightarrow \Phi_a^\dagger\, \overline{\ell}_\alpha)}{\sum\limits_{\beta=e,\mu, \tau} \sum\limits_{b=1}^{n_H} \left[ \Gamma(N_i \rightarrow \Phi_b \ell_\beta) + \Gamma(N_i \rightarrow \Phi_b^\dagger\, \overline{\ell}_\beta)\right]} \; .
\end{align}
Note that $\Gamma(N_i \rightarrow \Phi_a \ell_\alpha)$ and $\Gamma(N_i \rightarrow \Phi_a^\dagger\, \overline{\ell}_\alpha)$ are the heavy neutrino decays into leptons and antileptons of flavour $\alpha$, respectively. Summing $\varepsilon_{i \alpha}^a$ for all $\alpha$ and $a$, the total (unflavoured) CP asymmetry $\varepsilon_i$ in the $N_i$ decays is obtained,
\begin{equation}
    \varepsilon_i = \sum_{\alpha=e,\mu,\tau} \sum_{a=1}^{n_H} \varepsilon_{i \alpha}^a \; .
    \label{eq:cptotunflavoured}
\end{equation}
In the usual type-I leptogenesis scenario, the CP asymmetries are generated through the interference between the tree level and one-loop contributions of diagrams (\subref{fig:treetypeI})-(\subref{fig:wavetypeI}) in figure~\ref{fig:usualcontribution_CPasym}. At tree level the $N_i \rightarrow \Phi_a \ell_\alpha $ decay width is given by
\begin{align}
\Gamma(N_i \rightarrow \Phi_a \ell_\alpha) = \Gamma(N_i \rightarrow \Phi_a^\dagger\, \overline{\ell}_\alpha) = M_i \frac{\mathbf{H}^a_{i i}}{16 \pi} \; ,
\label{eq:Decaytree}
\end{align}
where $\mathbf{H}^a$ has been defined in eq.~\eqref{eq:YHdef}. The leading-order non-vanishing contributions to the CP asymmetries arising from the aforementioned interference is~\cite{Covi:1996wh,Branco:2011zb}

\begin{align}
    \varepsilon_{i \alpha}^a(\text{type-I}) &= \frac{1}{8 \pi \sum\limits_{a=1}^{n_H} \mathbf{H}^a_{i i}}  \sum_{b=1}^{n_H} \bigg\{\sum_{j=1}^{n_R} \sum_{\beta=e, \mu, \tau} \text{Im}\left[\mathbf{Y}^{a \ast}_{\alpha i} \mathbf{Y}^{b \ast}_{\beta i} \mathbf{Y}^{b}_{\alpha j} \mathbf{Y}^{a}_{\beta j}\right] f(r_{j i}) \nonumber \\ &+\sum_{j\neq i=1}^{n_R} \text{Im}\left[\mathbf{Y}^{a \ast}_{\alpha i} \mathbf{H}^b_{i j} \mathbf{Y}^{a}_{\alpha j} \right] g(r_{j i}) +\sum_{j\neq i=1}^{n_R} \text{Im}\left[ \mathbf{Y}^{a \ast}_{\alpha i} \mathbf{H}^b_{j i} \mathbf{Y}^{a}_{\alpha j} \right] g^\prime(r_{j i}) \bigg\} \; ,
    \label{eq:CPasymtypeI}
\end{align}
where $r_{j i} = M_j^2/M_i^2$. The loop functions $f(x)$ and $g(x)$ correspond to the vertex correction [diagram (\subref{fig:vertextypeI}) in figure~\ref{fig:usualcontribution_CPasym}] and $g^\prime(x)$ to the self-energy one [diagrams (\subref{fig:wavetypeI}) in figure~\ref{fig:usualcontribution_CPasym}]. These functions are given by
\begin{align}
f(x) = \sqrt{x} \left[1-(1+x) \log\left(1 + \frac{1}{x}\right) \right] \; , \; g(x) = \sqrt{x} g^\prime(x)= \frac{\sqrt{x}}{1-x} \, .
\label{eq:loopI}
\end{align}

Due to the presence of the scalars $h_k$, which couple to a pair of RH neutrino fields and to $\Phi^\dag_a \Phi_b$ [see eqs.~\eqref{eq:Deltadef} and~\eqref{eq:muCPdef}, respectively], there will be additional contributions to the one-loop $N_i \rightarrow \Phi_a \ell_\alpha$ decay diagrams, as depicted in figure~\ref{fig:newcontribution_CPasym}. Moreover, the interference of new $N_i \rightarrow \ell_\alpha h_k \Phi_a$ three-body decay diagrams, presented in figure~\ref{fig:newcontributiontree_CPasym}, must also be taken into account. These new contributions to the CP asymmetry were computed in ref.~\cite{LeDall:2014too} for a single Higgs doublet and a real scalar singlet. In this work, we present the results for the general case of $n_R$ RH neutrino fields, $n_H$ Higgs doublets and $n_S$ complex scalar singlets. For the diagrammatic computations we have used the standard Majorana Feynman rules~\cite{Gluza:1991wj,Denner:1992vza,Denner:1992me}. 

\subsection{New singlet-induced contributions to the CP asymmetry}
\label{sec:asymmetry}

\begin{figure}[t!]
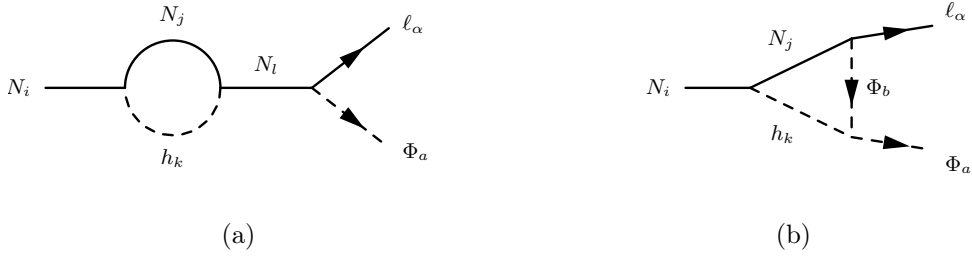

    \centering
     \begin{subfigure}[b]{0.4\textwidth}
         \centering
         \includegraphics[scale=0.9,trim={11.5cm 8.5cm 2.5cm 16cm},clip]{diagrams_CPasymmetry.pdf}
         \caption{}
         \label{fig:newwave}
     \end{subfigure}
     \hspace{1cm}
      \begin{subfigure}[b]{0.4\textwidth}
         \centering
         \includegraphics[scale=0.9,trim={4.5cm 8.5cm 11.5cm 16cm},clip]{diagrams_CPasymmetry.pdf}
         \caption{}
         \label{fig:newvertex}
     \end{subfigure}\\
    \caption{New one-loop diagrams contributing to the CP asymmetry in $N_i \rightarrow \ell_\alpha \Phi_a$ decays via the interference with the corresponding tree-level diagram (a) of figure~\ref{fig:usualcontribution_CPasym}. (a) Wavefunction contribution $\sim \mathcal{O}(Y^2 \Delta^2)$. (b) Vertex contribution $\sim \mathcal{O}(Y^2\Delta \mu)$.}
    \label{fig:newcontribution_CPasym}
\end{figure}

\begin{figure}[t!]
    \centering
    \includegraphics[scale=0.9,trim={4.5cm 12cm 2.5cm 12cm},clip]{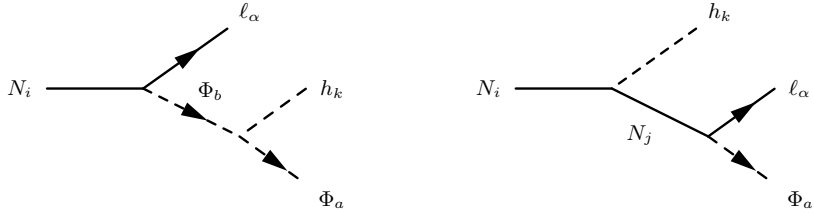}
    \caption{Interfering $N_i \rightarrow \ell_\alpha h_k \Phi_a$ tree-level 3-body decays. Left [right] diagram is of order~$\sim \mathcal{O}(Y \mu)$ [$\mathcal{O}(Y \Delta)$] in the couplings.  }
    \label{fig:newcontributiontree_CPasym}
\end{figure}

We start with the new $h_k$-mediated wavefunction contribution to the CP asymmetry $\varepsilon_{i \alpha}^a$ in the $N_i \rightarrow \Phi_a \ell_\alpha$ decays, for which the relevant diagrams are labeled as~(\subref{fig:newwave}) in figure~\ref{fig:newcontribution_CPasym}. Denoting generically the couplings in eqs.~\eqref{eq:YHdef}-\eqref{eq:muCPdef} by $Y$, $\Delta$ and $\mu$, it is straightforward to see that these diagrams scale as $Y^2\Delta^2$. Their interference with the corresponding tree-level ones leads to the following wavefunction contribution to the CP asymmetry $\varepsilon_{i \alpha}^a$ at one-loop level:
\begin{align}
    \varepsilon_{i \alpha}^a(\text{wave})& = \frac{1}{8 \pi \sum\limits_{a=1}^{n_H}  \mathbf{H}^a_{i i}}\sum_{j,l \neq i = 1}^{n_R} \sum_{k=1}^{2 n_S} (1+\delta_{j l}) \bigg\{\text{Im}\left[ \mathbf{Y}_{\alpha l}^{a \ast} \mathbf{\Delta}_{l j}^k \mathbf{\Delta}_{j i}^{k \ast} \mathbf{Y}_{\alpha i}^{a} \right] \mathcal{F}_{\text{w}, L L}^{i j k l} \nonumber \\ 
    & + \text{Im}\left[\mathbf{Y}_{\alpha l}^{a \ast} \mathbf{\Delta}_{l j}^{k \ast} \mathbf{\Delta}_{j i}^{k \ast} \mathbf{Y}_{\alpha i}^{a} \right] \mathcal{F}_{\text{w}, L R}^{i j k l} 
    + \text{Im}\left[ \mathbf{Y}_{\alpha l}^{a \ast} \mathbf{\Delta}_{l j}^k \mathbf{\Delta}_{j i}^k \mathbf{Y}_{\alpha i}^{a} \right] \mathcal{F}_{\text{w}, R L}^{i j k l} \nonumber \\
    & + \text{Im}\left[\mathbf{Y}_{\alpha l}^{a \ast} \mathbf{\Delta}_{l j}^{k \ast} \mathbf{\Delta}_{j i}^k \mathbf{Y}_{\alpha i}^{a} \right] \mathcal{F}_{\text{w}, R R}^{i j k l}\bigg\} \; ,
    \label{eq:CPasymwave}
\end{align}
being the loop functions given by,
\begin{align}
\mathcal{F}_{\text{w}, L L}^{i j k l} &= \frac{\sqrt{\rho_{i j k}} \sqrt{\rho_{i j k} + 4 r_{j i}}}{2 (1-r_{l i})}  \, , \;  \mathcal{F}_{\text{w}, L R}^{i j k l} =  \frac{\sqrt{\rho_{i j k}} \sqrt{r_{j i}} \sqrt{r_{l i}}}{1-r_{l i}}\, , \nonumber \\
\mathcal{F}_{\text{w}, R L}^{i j k l} &=  \frac{\sqrt{\rho_{i j k}} \sqrt{r_{j i}}}{1-r_{l i}} \, , \; \mathcal{F}_{\text{w}, R R}^{i j k l} = \frac{\sqrt{\rho_{i j k}} \sqrt{r_{l i}} \sqrt{\rho_{i j k} + 4 r_{j i}}}{2 (1-r_{l i})} \, ,
\label{eq:loopwave}
\end{align}
where $\sigma_{k i} = m_{h_k}^2/M_i^2$ and $\rho_{i j k} = (1-r_{j i}-\sigma_{k i})^2 - 4 r_{ji} \sigma_{ki}$. Note that, in order for the CP-asymmetries induced by the $h_k$ scalars to be non-zero one must guarantee that the kinematic constraint $ M_i > M_j + m_{h_k} $, i.e.~$  \sqrt{r_{j i}} + \sqrt{\sigma_{k i}} <1 $, is verified. In fact, the one-loop and the tree-level diagrams depicted in figures~\ref{fig:newcontribution_CPasym} and~\ref{fig:newcontributiontree_CPasym}, respectively, only lead to a non-vanishing imaginary part for the CP-asymmetry if the latter kinematic constraint is fulfilled. Hence, the decay $N_i \rightarrow N_j h_k$ needs to be allowed.

The $h_k \Phi^\dag_a \Phi_b$ couplings of eq.~\eqref{eq:muCPdef} will generate new one-loop vertex contributions to~$\varepsilon_{i \alpha}^a$ stemming from diagrams~(\subref{fig:newvertex}) of figure~\ref{fig:newcontribution_CPasym}, which are of order $\mathcal{O}(Y^2\Delta \mu)$. For this case we obtain
\begin{align}
    \varepsilon_{i \alpha}^a(\text{vertex}) &= \frac{1}{8 \pi M_i \sum\limits_{a=1}^{n_H}  \mathbf{H}^a_{i i}}\sum_{j\neq i = 1}^{n_R} \sum_{k=1}^{2 n_S} \sum_{b=1}^{n_H} \bigg\{\text{Im}\left[ \mathbf{Y}_{\alpha i}^{a } \mathbf{Y}_{\alpha j}^{b\ast} \mathbf{\Delta}_{i j}^k \tilde{\mu}_{a b, k} \right] \mathcal{F}_{\text{v}, L L}^{i j k} \nonumber \\
    &+ \text{Im}\left[ \mathbf{Y}_{\alpha i}^{a } \mathbf{Y}_{\alpha j}^{b\ast} \mathbf{\Delta}_{i j}^{k \ast} \tilde{\mu}_{a b, k} \right] \mathcal{F}_{\text{v}, R L}^{i j k} \bigg\} \; ,
    \label{eq:CPasymvertex}
\end{align}
where the loop functions read
\begin{align}
\mathcal{F}_{\text{v}, L L}^{i j k} &= - \sqrt{\rho_{i j k}} + r_{j i} \log \left[ \frac{\sqrt{\rho_{i j k} + 4 r_{j i} \sigma_{k i}} - \sqrt{\rho_{i j k}}}{\sqrt{\rho_{i j k} + 4 r_{j i} \sigma_{k i}} + \sqrt{\rho_{i j k}}} \right] \, , \nonumber \\
\mathcal{F}_{\text{v}, R L}^{i j k} &= \sqrt{r_{j i}} \log \left[ \frac{\sqrt{\rho_{i j k} + 4 r_{j i} \sigma_{k i}} - \sqrt{\rho_{i j k}}}{\sqrt{\rho_{i j k} + 4 r_{j i} \sigma_{k i}} + \sqrt{\rho_{i j k}}} \right] \, .
\label{eq:loopvertex}
\end{align}

The interference between the 3-body decay diagrams of figure~\ref{fig:newcontributiontree_CPasym} gives corrections to $\varepsilon_\text{CP}$ of the same order of the vertex contribution, i.e. of order $\mathcal{O}(Y^2 \Delta \mu)$. For this case the CP asymmetry is computed as
\begin{equation}
    \varepsilon_{i \alpha}^a(\text{3-body decay}) \simeq \frac{\sum\limits_{k=1}^{2n_S} \left[\Gamma(N_i \rightarrow \Phi_a \ell_\alpha h_k) - \Gamma(N_i \rightarrow \Phi_a^\dagger \overline{\ell}_\alpha h_k)\right]}{\sum\limits_{\beta=e,\mu,\tau} \sum\limits_{b=1}^{n_H} \left[ \Gamma(N_i \rightarrow \Phi_b \ell_\beta) + \Gamma(N_i \rightarrow \Phi_b^\dagger \overline{\ell}_\beta)\right]} \; ,
\end{equation}
where we neglected the three-body decay rate in the denominator since it is subdominant compared to the two-body decay, due to its reduced phase space. We obtain
\begin{align}
    \varepsilon_{i \alpha}^a(\text{3-body decay}) &=\frac{1}{8 \pi M_i \sum\limits_{a=1}^{n_H} \mathbf{H}^a_{i i}}\sum_{j\neq i = 1}^{n_R} \sum_{k=1}^{2 n_S} \sum_{b=1}^{n_H}\bigg\{\text{Im}\left[ \mathbf{Y}_{\alpha i}^{b \ast} \mathbf{Y}_{\alpha j}^a \mathbf{\Delta}_{i j}^k \tilde{\mu}_{a b, k} \right] \mathcal{F}_{\text{3BD}, L L}^{i j k} \nonumber \\
    &+ \text{Im}\left[ \mathbf{Y}_{\alpha i}^{b \ast} \mathbf{Y}_{\alpha j}^a \mathbf{\Delta}_{i j}^{k \ast} \tilde{\mu}_{a b, k} \right] \mathcal{F}_{\text{3BD}, R L}^{i j k} \bigg\} \; ,
    \label{eq:CPasym3body}
\end{align}
where,
\begin{align}
\mathcal{F}_{\text{3BD}, L L}^{i j k} &=  - \sqrt{\rho_{i j k}r_{j i}} + \sqrt{r_{j i}}\log \left[ \frac{\sqrt{\rho_{i j k} + 4 r_{j i} \sigma_{k i}} + 2\sigma _{ki}+ \sqrt{\rho_{i j k}}}{\sqrt{\rho_{i j k} + 4 r_{j i} \sigma_{k i}} + 2\sigma _{ki} - \sqrt{\rho_{i j k}}} \right] \, , \nonumber \\
\mathcal{F}_{\text{3BD}, R L}^{i j k} &= r_{j i} \log \left[ \frac{\sqrt{\rho_{i j k} + 4 r_{j i} \sigma_{k i}} + 2\sigma _{ki}+ \sqrt{\rho_{i j k}}}{\sqrt{\rho_{i j k} + 4 r_{j i} \sigma_{k i}} + 2\sigma _{ki} - \sqrt{\rho_{i j k}}} \right] \, .
\label{eq:loop3body}
\end{align}

Combining the different contributions shown above leads to
\begin{equation}
   \varepsilon_{i \alpha}^a = \varepsilon_{i \alpha}^a(\text{type-I}) + \varepsilon_{i \alpha}^a(\text{wave}) + \varepsilon_{i \alpha}^a(\text{vertex}) + \varepsilon_{i \alpha}^a(\text{3-body decay}) \; .
   \label{eq:fullCP}
\end{equation}
These results are consistent with the ones obtained in ref.~\cite{LeDall:2014too} in the limit of a single Higgs doublet and one real scalar singlet, apart from the 3-body decay contribution in eq.~\eqref{eq:CPasym3body} where the `$\text{Im}[ \cdots]$' coefficients in front of the $LL$ and $RL$ functions are exchanged.

As mentioned before, we are interested in imposing CP at the Lagrangian level and breaking it spontaneously through non-zero complex VEVs acquired by the singlets $S_k$ above the leptogenesis scale. Within this scenario, a few remarks can be made regarding the link between vacuum CP and the CP asymmetry:
\begin{itemize}
    \item By imposing CP conservation, the Yukawa couplings $\YD$, $\YR$, $\YR^\prime$, the bare mass term $\MR^0$ of eq.~\eqref{eq:LYuk} and the scalar potential potential parameters are real. Consequently, the unitary matrix $\mathbf{V}_L$ of eq.~\eqref{eq:VLmix} is orthogonal and $\tilde{\mu}_{a b, k}$ of eq.~\eqref{eq:muCPdef} is real. Hence, only the matrix entries of $\mathbf{Y}^a$ and $\mathbf{\Delta}^k$, given by eqs.~\eqref{eq:YHdef} and~\eqref{eq:Deltadef}, respectively, can be complex. If this is the case, the `$\text{Im}[\cdots]$' coefficients in the CP asymmetries do not vanish in general. More specifically, the rotation $\Ur$ provides the only connection to the CP violation encoded in the scalar complex VEVs, as explained in section~\ref{sec:framework}.
    \item Focusing on the case where $\Ur$ is complex, i.e. when SCPV is successfully communicated to the heavy neutrino sector, $\mathbf{Y}^a$ and $\mathbf{\Delta}^k$ are a priori general complex matrices. Hence, one expects that the different contributions to the total $\varepsilon_{\text{CP}}$ of eq.~\eqref{eq:fullCP} are non-zero. Interestingly, since the type-I CP-asymmetry contribution of eq.~\eqref{eq:CPasymtypeI} depends on the product of four $\mathbf{Y}^a$ matrix elements, one can derive conditions such that this contribution vanishes.
    \item In the scenario where $\varepsilon(\text{type-I})=0$, one would generate the high-energy CP violation needed to explain the BAU entirely through the new singlet-assisted diagrams. To achieve this, one needs to guarantee that $\mathbf{Y}^{a}$ have some special properties. Namely, if the elements of a given row of the $\mathbf{Y}^a$ matrix have the same phase, i.e. $\arg[\mathbf{Y}^{a}_{\alpha i}] = \arg[\mathbf{Y}^{a}_{\alpha j}]$ for $i \neq j$ and for all $j$, the Hermitian matrix $\mathbf{H}^a$ is necessarily real and therefore the second and third `$\text{Im}[\cdots]$' terms in eq.~\eqref{eq:CPasymtypeI} for $\varepsilon^a_{i \alpha}(\text{type-I})$ vanish. In such case, summing over all lepton flavours leads to a vanishing unflavoured CP asymmetry, i.e. $\varepsilon^a_{i}(\text{type-I})=0$. In order for the flavoured type-I CP asymmetry to vanish one needs to verify that the elements of a given row of the $\mathbf{Y}^a$ and $\mathbf{Y}^b$ matrices have the same phase, i.e. $\arg[\mathbf{Y}^{a}_{\alpha i}] = \arg[\mathbf{Y}^{b}_{\alpha j}]$ for $i \neq j$ and for all $j$ and $b$. Such specific properties of the Yukawa matrices $\mathbf{Y}^{a}$ may result, e.g., from a flavour symmetry.
\end{itemize}
In section~\ref{sec:model} we will present a very simple flavour model in which the new CP asymmetries induced by the scalar states $h_k$ provide the only link between the vacuum CP phases and the BAU.

\subsection{Unflavoured Boltzmann equations}
\label{sec:BEs}

\begin{figure}[t!]
    \centering
     \begin{subfigure}[b]{0.45\textwidth}
         \centering
         \includegraphics[scale=0.9]{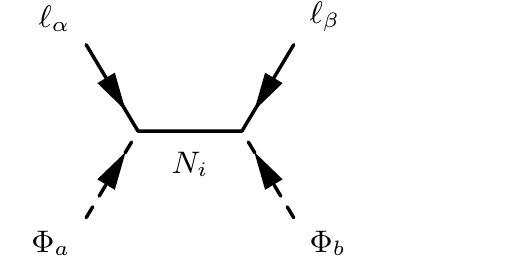}
         \caption{ }
         \label{fig:RIS1}
     \end{subfigure}
     \hspace{+1cm}
      \begin{subfigure}[b]{0.45\textwidth}
         \centering
         \includegraphics[scale=0.9]{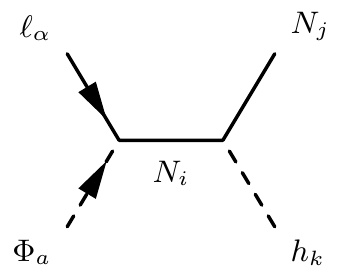}
         \caption{}
         \label{fig:RIS2}
     \end{subfigure}\\
     \vspace{+0.5cm}
     \begin{subfigure}[b]{0.45\textwidth}
         \centering
         \includegraphics[scale=0.9]{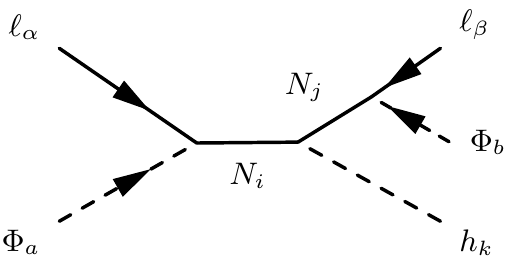}
         \caption{}
         \label{fig:RIS3}
     \end{subfigure}
     \hspace{+1cm}
      \begin{subfigure}[b]{0.45\textwidth}
         \centering
         \includegraphics[scale=0.9]{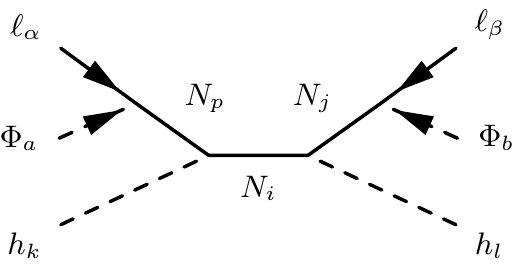}
        \caption{}
         \label{fig:RIS4}
     \end{subfigure}
    \caption{Scattering diagrams containing {\em real intermediate states}~(RIS). These must be subtracted in order to avoid inconsistent source terms in the BEs (see text for details). (a) Usual $\Delta L = 2$ scattering $\ell_\alpha \Phi_a \leftrightarrow \overline{\ell}_\beta \Phi_b^\dagger$ $s$-channel contribution $\sim \mathcal{O}(Y^4)$. (b) New $\Delta L = 1$ scattering $\ell_\alpha \Phi_a \leftrightarrow N_1 h_k$ $s$-channel contribution $\sim \mathcal{O}(Y^2 \Delta^2)$. (c) New $\Delta L = 2$ scattering $\ell_\alpha \Phi_a \leftrightarrow \overline{\ell}_\beta \Phi_b^\dagger h_k$ RIS contribution $\sim \mathcal{O}(Y^4 \Delta^2)$. (d) New $\Delta L = 2$ scattering $\ell_\alpha \Phi_a h_k \leftrightarrow \overline{\ell}_\beta \Phi_b^\dagger h_l$ RIS contribution $\sim \mathcal{O}(Y^4 \Delta^4)$.}
    \label{fig:diagsRIS}
\end{figure}

The lepton (and baryon) asymmetry produced through leptogenesis can be computed by solving the BEs that describe the out-of-equilibrium dynamics of the various processes involving the heavy Majorana neutrinos $N_i$. In this section, we derive the system of (classical) BEs relevant for the type of models we are interested in (general aspects related to BEs are reviewed in appendix~\ref{sec:genBEs}). Before presenting our results, the following comments are in order:  
\begin{itemize}
    \item We will restrict our analysis to the temperature regime $T > 10^{12}$~GeV, i.e. the unflavoured scenario where the CP and lepton asymmetries are summed over all flavours (for reviews on flavour effects in leptogenesis see refs.~\cite{Abada:2006fw,Nardi:2006fx,Abada:2006ea,Blanchet:2006be,Dev:2017trv}). For the sake of generality, we will present the BEs for an arbitrary number of RH neutrinos $N_{i}$ ($i=1, \cdots, n_R$), where the masses are ordered as $M_1 < M_2 < ... <M_{n_R}$, being $N_1$ the lightest heavy neutrino. We do not consider in this work thermal corrections to the masses and to the CP asymmetries (for an analysis on this subject the reader is addressed to ref.~\cite{Giudice:2003jh}).
    \item In the standard type-I seesaw leptogenesis, the CP asymmetries needed to generate the BAU are of order~$\mathcal{O}(Y^4)$ in the Yukawa couplings (see figure~\ref{fig:usualcontribution_CPasym}). Hence, in order to obtain BEs respecting the Sakharov conditions~\cite{Sakharov:1967dj}, one needs to take into account $\Delta L = 2$ scattering processes up to order $\mathcal{O}(Y^4)$. The corresponding diagrams exhibit what is known as {\em real intermediate states}~(RIS) that must be subtracted in order to obtain consistent BEs~\cite{Kolb:1979qa,Buchmuller:1997yu,Buchmuller:2000nq}.\footnote{Without such a procedure a $(B-L)$-asymmetry could be generated via source terms in thermal equilibrium, which is forbidden by the CPT symmetry~\cite{Dimopoulos:1978kv,Kolb:1979qa,Dolgov:1981hv}.} This is the case of the $s$-channel $N_i$ mediated $\Delta L = 2$ scattering  $\ell_\alpha \Phi_a \leftrightarrow \overline{\ell}_\beta \Phi_b^\dagger$ shown in diagram (\subref{fig:RIS1}) of figure~\ref{fig:diagsRIS}. 
    It has a RIS, since the mediating neutrino state can be produced on-shell, i.e. $\ell_\alpha \Phi_a \leftrightarrow N_i \leftrightarrow \overline{\ell}_\beta \Phi_b^\dagger$. Hence, this intermediate decay must be subtracted since it is already accounted for in the BEs.
    \item
    The presence of additional scalar mass-eigenstates coupling to heavy Majorana neutrinos opens up a new decay channel if $M_i>M_j + m_{h_k}$, namely $N_i \rightarrow N_j h_k$. The total decay widths for $N_i$ are
           \begin{align}
           \Gamma_i &= \sum_{a=1}^{n_H} \sum_{\alpha= e, \mu,\tau} \left[\Gamma(N_i \rightarrow \Phi_a \ell_\alpha) + \Gamma(N_i \rightarrow \Phi_a^\dagger \overline{\ell}_\alpha) \right]+ \sum_{j=1}^{n_R} \sum_{k=1}^{2 n_S} \Gamma(N_i \rightarrow N_j h_k) \; ,
           \label{eq:neutrinototal}
           \end{align}
          where the expressions for the decay rates are given in eqs.~\eqref{eq:Decaytree} and~\eqref{eq:newdecayG}. Due to these extra interactions, compared to standard type-I seesaw framework, one needs to consider $\Delta L = 2$ scattering processes up to order $\mathcal{O}(Y^4 \Delta^4)$ and subtract the appropriate RIS to achieve consistent BEs. In fact, as shown in figure~\ref{fig:diagsRIS}, diagram~(\subref{fig:RIS2}) for the $\Delta L = 1$ scattering $\ell_\alpha \Phi_a \leftrightarrow N_j h_k$ process has a RIS corresponding to its $N_i$ ($i \neq j$) mediated $s$-channel. The $\Delta L = 2$ two-to-three body scattering $\ell_\alpha \Phi_a \leftrightarrow \overline{\ell}_\beta \Phi_b^\dagger h_k$ in figure~\ref{fig:diagsRIS}(\subref{fig:RIS3}) and three-to-three scattering $\ell_\alpha \Phi_a h_k \leftrightarrow \overline{\ell}_\beta \Phi_b^\dagger h_l$ in figure~\ref{fig:diagsRIS}(\subref{fig:RIS4}), both contain RIS diagrams. By systematically subtracting these RIS contributions one reaches a set of coupled BEs meeting all the Sakharov conditions with no inconsistencies. The procedure outlined here was performed in detail in ref.~\cite{LeDall:2014too} for the 2RH neutrino case ($n_R=2$), and we also followed the approach of ref.~\cite{Giudice:2003jh}. Furthermore, there are additional $\Delta L = 1$ scatterings which contain $t$-channel RIS (see details in appendix~\ref{sec:scatterings}).

    \item For the purposes of our work, we will neglect the three-body decay reaction densities in the BEs since these are subdominant when compared to usual two-body decays. Furthermore, the two-to-three and three-to-three scatterings are also subdominant when compared to the two-to-two scattering processes, due to their reduced phase space. Also, we will not consider the $\Delta L=2$ scattering contributions in the BEs.
    
\end{itemize}
Taking into account the general framework outlined above, the set of BEs for leptogenesis is given by~(for details see appendix~\ref{sec:genBEs})
\begin{align}
    \frac{d N_{N_i}}{d z} & = -(D_i + S_i) (N_{N_i} - N_{N_i}^{\text{eq}}) - \sum_{j=1}^{n_R} S_{i j} (N_{N_i} N_{N_j} - N_{N_i}^{\text{eq}} N_{N_j}^{\text{eq}}) \label{eq:BEsNi}  \\
    & + \sum_{j=1}^{n_R} \left[ \left(\frac{N_{N_i}^{\text{eq}}}{N_{N_j}^{\text{eq}}} D_{i j} + D_{j i}\right) (N_{N_j} - N_{N_j}^{\text{eq}}) - \left(D_{i j} + \frac{N_{N_j}^{\text{eq}}}{N_{N_i}^{\text{eq}}} D_{j i}\right) (N_{N_i} - N_{N_i}^{\text{eq}}) \right] \; , \nonumber \\
    \frac{d N_{B-L}}{d z} & = - \sum_{i=1}^{n_R} \varepsilon_i D_i (N_{N_i} - N_{N_i}^{\text{eq}}) - W N_{B-L} \;,\;\;\;\;N_{N_i}^{\text{eq}} = \frac{3}{8} z_i^2 \mathcal{K}_2(z_i) \,. \label{eq:BEsBL}
\end{align}
In the above equations, $z_i = M_i/T$ and we use the notation $z\equiv z_1$. The quantity $N_{N_i}$ ($N_{N_j}^{\text{eq}}$) is the (equilibrium) $N_i$ number density. The temperature-dependent quantities $D_i(z)$, $D_{j i}(z)$, $S_i(z)$, $S_{i j}(z)$ and $W(z)$ are, respectively, the decay, the scattering and the washout terms. In particular, the coefficient $D_{i j}$ vanishes if $N_i \rightarrow N_j h_k$ is not kinematically allowed, i.e. for~$i \leq j$. Note that $\varepsilon_i$ is the $N_i$ unflavoured CP asymmetry computed via eqs.~\eqref{eq:cptotunflavoured} and~\eqref{eq:fullCP}. The above system is solved in order to compute~$N_{B-L}$ and determine the BAU using eq.~\eqref{eq:etaB}. We will take as initial conditions $N_{N_i}(z=0) = N_{N_i}^{\text{eq}}(z=0) = 3/4$ and $N_{B-L}(z=0) = 0$. The latter corresponds to the case of a Universe with no initial $(B-L)$-asymmetry. As explained above, by including all necessary diagrams shown in figure~\ref{fig:diagsRIS}, we obtain BEs equations meeting all Sakharov conditions required to successfully generate an asymmetry from an initially symmetric state. Namely, the lepton-asymmetry production in the term~$D_i$, the CP violation in $\varepsilon_i$, and the departure from thermal equilibrium of $N_i$ through the term~$N_{N_i}-N_{N_i}^{\text{eq}}$. If any of these terms vanishes, a $(B-L)$-asymmetry cannot be generated. 

We now turn our attention to the specific formulae for the BE coefficients.
\begin{itemize}

    \item \textbf{Decays:}
    
    The expressions for the decay parameters $D_i(z)$ and $D_{i j}(z)$ are [see eq.~\eqref{eq:Decaygen}]
    \begin{align}
    D_i(z) = K_i r_{i 1} z \frac{\mathcal{K}_1(z_i)}{\mathcal{K}_2(z_i)} \; , \;D_{i j}(z) =
        K_{i j} r_{i 1} z \frac{\mathcal{K}_1(z_j)}{\mathcal{K}_2(z_j)}\, ,
    \label{eq:decayparam}
    \end{align}
    with
    \begin{align}
    K_i = \sum_{\alpha=e, \mu, \tau} \sum_{a=1}^{n_H} \frac{\Gamma(N_i \rightarrow \Phi_a \ell_\alpha) + \Gamma(N_i \rightarrow \Phi_a^\dagger \overline{\ell}_\alpha)}{H(T=M_i)} , K_{i j} = \sum_{k=1}^{2 n_S} \frac{\Gamma(N_i \rightarrow N_j h_k)}{H(T=M_i)} , 
    \end{align}
    where the Hubble parameter is given by eq.~\eqref{eq:Hubble} and the expression for the LNV decay entering $K_i$ is the one of eq.~\eqref{eq:Decaytree}. Note that the LNV inverse decay must be taken into account while computing the washout parameter $W(z)$. Furthermore, the decay rate for the tree-level $N_i \rightarrow N_j h_k$ process is
    \begin{align}
    \Gamma(N_i \rightarrow N_j h_k) = \frac{M_i \sqrt{\rho_{j i k}}}{16 \pi} \left\{ (1 + r_{j i} - \sigma_{k i}) \left|\mathbf{\Delta}_{i j}^k\right|^2 + 2 \sqrt{r_{j i}} \text{Re}\left[ (\mathbf{\Delta}^k_{i j})^2 \right]\right\}\; ,
    \label{eq:newdecayG}
    \end{align}
    being kinematically allowed only for $M_i > M_j + m_{h_k}$. For the 2RH neutrino case, the result obtained in ref.~\cite{LeDall:2014too} is recovered taking $i=2$ and $j=1$.
    
\begin{figure}[t!]
  \centering
     \begin{subfigure}[b]{0.9\textwidth}
        \centering
         \includegraphics[scale=0.95,trim={4cm 23.5cm 2.5cm 1cm},clip]{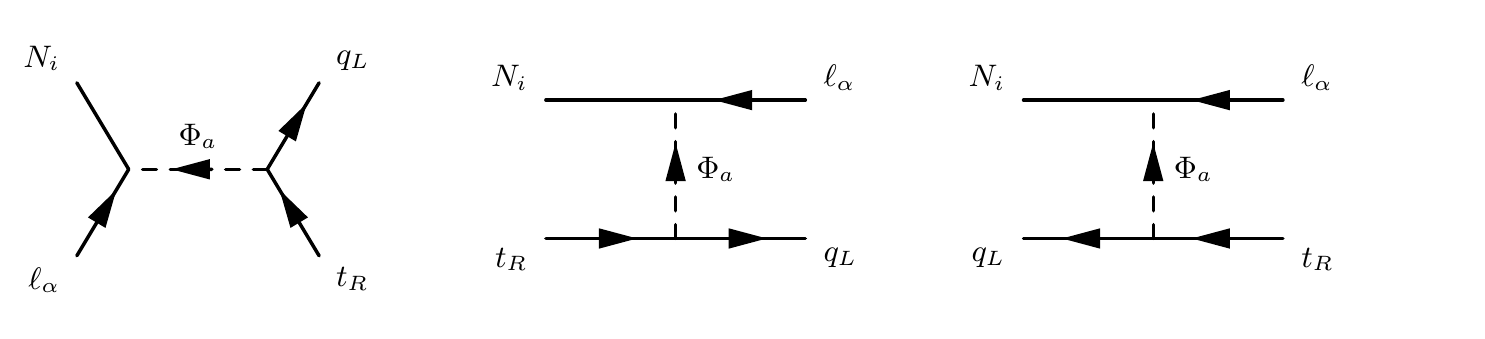}
         \caption{Usual type-I seesaw contributions: $N_i\ell_\alpha\leftrightarrow q_L t_R$, $N_it_R\leftrightarrow \ell_\alpha q_L $ and $N_iq_L\leftrightarrow \ell_\alpha t_R $.}
        \label{fig:DeltaLeq1_usual}
    \end{subfigure}\\
    \begin{subfigure}[b]{0.9\textwidth}
        \centering
       \includegraphics[scale=0.95,trim={4cm 20.1cm 2.5cm 4cm},clip]{diagramsDeltaLeq1.pdf}
        \caption{$N_i\ell_\alpha\leftrightarrow\Phi_a h_k$}
         \label{fig:DeltaLeq1_Nltophih}
    \end{subfigure}
    \\
   \begin{subfigure}[b]{0.9\textwidth}
        \centering
        \includegraphics[scale=0.95,trim={4cm 16.9cm 2.5cm 7.5cm},clip]{diagramsDeltaLeq1.pdf}
      \caption{$N_ih_k\leftrightarrow\ell_\alpha \Phi_a$}
        \label{fig:DeltaLeq1_Nhtophil}     
\end{subfigure}\\
    \begin{subfigure}[b]{0.9\textwidth}
\centering
       \includegraphics[scale=0.95,trim={4cm 13.9cm 2.5cm 11cm},clip]{diagramsDeltaLeq1.pdf}
        \caption{$N_i\Phi_a\leftrightarrow\ell_\alpha h_k$}
         \label{fig:DeltaLeq1_Nphitohl}
     \end{subfigure}
    \caption{Scattering contributions with $\Delta L=1$.}
    \label{fig:DeltaLeq1_scattering}
\end{figure}

\item \textbf{Scatterings:}  
 
The Feynman diagrams for all $\Delta L=1$ two-body scattering processes included in our analysis are shown in figure~\ref{fig:DeltaLeq1_scattering}, with the corresponding reduced cross sections given in appendix~\ref{sec:scatterings}. Diagrams (\subref{fig:DeltaLeq1_usual}) are the usual ones of standard type-I seesaw leptogenesis, namely the $s$-channel $N_i\ell_\alpha\leftrightarrow q_L t_R$ and $t$-channel $N_it_R\leftrightarrow \ell_\alpha q_L$ and $N_iq_L\leftrightarrow \ell_\alpha t_R $~\cite{Buchmuller:2004nz}. Due to the presence of the new scalar particles $h_k$, new $\Delta L=1$ scatterings are allowed: (\subref{fig:DeltaLeq1_Nltophih}) $N_i\ell_\alpha\leftrightarrow\Phi_a h_k$, (\subref{fig:DeltaLeq1_Nhtophil}) $N_ih_k\leftrightarrow\ell_\alpha \Phi_a$ and (\subref{fig:DeltaLeq1_Nphitohl}) $N_i\Phi_a\leftrightarrow\ell_\alpha h_k$, with (\subref{fig:DeltaLeq1_Nltophih}) and (\subref{fig:DeltaLeq1_Nhtophil}) having an $s-$ and $t$-channel diagram, while (\subref{fig:DeltaLeq1_Nphitohl}) only occurs via $t$-channel mediation. As mentioned before, the $s$-channel $N_i\ell_\alpha\leftrightarrow\Phi_a h_k$ diagram features a RIS which must be subtracted to obtain a consistent set of BEs. Similarly, the $t$-channel Higgs mediated diagrams for $N_ih_k\leftrightarrow\ell_\alpha \Phi_a$ and $N_i\Phi_a\leftrightarrow\ell_\alpha h_k$ also contain RIS -- see appendix~\ref{sec:scatterings} for details on the subtraction procedure. In comparison to ref.~\cite{LeDall:2014too}, where only the dominant contributions to these processes were included, here we consider and compute all the tree-level contributions for the aforementioned processes. The scattering parameters $S_i$ entering the BEs in eqs.~\eqref{eq:BEsNi} and~\eqref{eq:BEsBL} are given by [see eqs.~\eqref{eq:equi} and~\eqref{eq:Scatteringgen}]
    \begin{align}
        S_i = \frac{\gamma_i^\text{eq}}{n_{N_i}^\text{eq} H(z) z} \; , \; n_{N_i}^\text{eq}=\frac{M_i^2 T}{2 \pi^2} \mathcal{K}_2(z_i) \; ,
    \end{align}
    where $H(z) \equiv H(T= M_1/z)$. The reaction density is defined as
    \begin{align}
        \gamma_i^\text{eq} &= \sum_{\alpha=e, \mu, \tau} \left[\gamma^\text{eq}(N_i \ell_{\alpha} \leftrightarrow q_L t_R) + \gamma^\text{eq}(N_i t_R  \leftrightarrow \ell_{\alpha} q_L) + \gamma^\text{eq}(N_i q_L \leftrightarrow \ell_{\alpha} t_R) \right]  \\
        & + \sum_{\alpha=e, \mu, \tau} \sum_{a=1}^{n_H} \sum_{k=1}^{2 n_S} \left[\gamma^\text{eq}(N_i \ell_{\alpha} \leftrightarrow \Phi_a h_k) + \gamma^\text{eq}(N_i h_k \leftrightarrow \ell_{\alpha} \Phi_a) + \gamma^\text{eq}(N_i \Phi_a \leftrightarrow \ell_{\alpha} h_k) \right] \; , \nonumber
    \end{align}
    which is computed through eq.~\eqref{eq:reacscattering} using the reduced cross sections presented in appendix~\ref{sec:scatterings} and taking into consideration all necessary RIS subtractions. It is worth stressing that the above LNV scattering processes will contribute to the washout coefficient~$W(z)$.

\begin{figure}[t!]
    \centering
     \begin{subfigure}[b]{0.9\textwidth}
         \centering
         \includegraphics[scale=0.95,trim={4cm 23.5cm 2.5cm 1cm},clip]{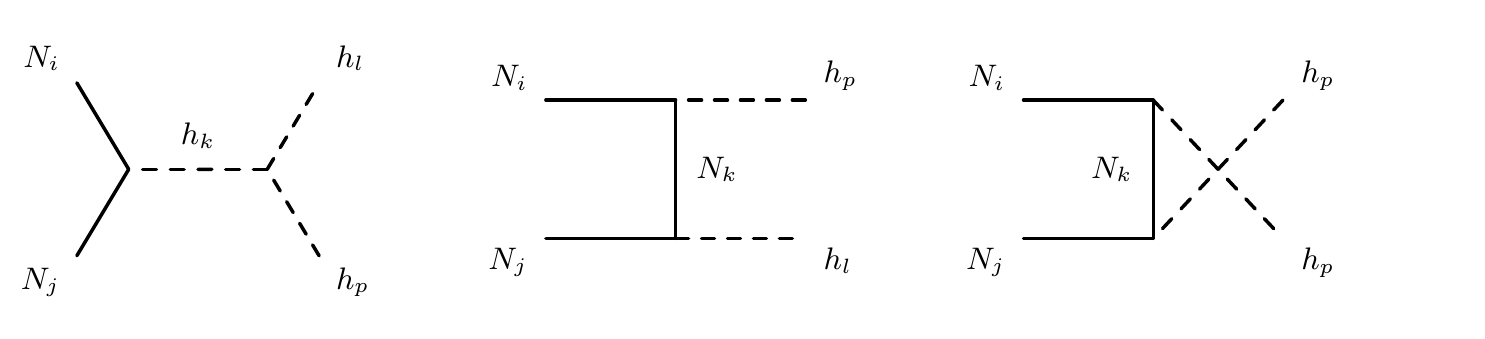}
        \caption{$N_i N_j \leftrightarrow h_p h_l$.}
         \label{fig:DeltaNeq2_NNtohh}
     \end{subfigure}\\
     \begin{subfigure}[b]{0.9\textwidth}
         \centering
        \includegraphics[scale=0.95,trim={4cm 20.5cm 2.5cm 4cm},clip]{diagramsDeltaNeq2.pdf}
        \caption{$N_i N_j \leftrightarrow \Phi_a \Phi_b$}
         \label{fig:DeltaNeq2_NNtohhphiphi}
    \end{subfigure}
    \caption{Neutrino annihilation scattering contributions with $\Delta N=2$.}
    \label{fig:DeltaNeq2_scattering}
\end{figure}

In figure~\ref{fig:DeltaNeq2_scattering} we present the novel heavy-neutrino annihilation diagrams which, not being LNV, are dubbed as $\Delta N = 2$ processes.\footnote{Such~$\Delta N = 2$ interactions appear, e.g. in the context of supersymmetric SO(10) unification~\cite{Plumacher:1996kc,Plumacher:1997ru,Plumacher:1998ex}.} Their existence is due to the interactions between the scalar singlets and heavy Majorana neutrinos, as well as to the triple scalar potential couplings involving $h_k$ -- see eqs~\eqref{eq:LYuk} and~\eqref{eq:muCPdef}, respectively. The process $N_i N_j \leftrightarrow h_p h_l$ shown in diagram (\subref{fig:DeltaNeq2_NNtohh}) was considered in ref.~\cite{AristizabalSierra:2014uzi} in the limit of a single heavy neutrino and one Yukawa coupling. The $s$-channel mediated diagram (\subref{fig:DeltaNeq2_NNtohhphiphi}) $N_i N_j \leftrightarrow \Phi_a \Phi_b$ was already computed in refs.~\cite{LeDall:2014too,Alanne:2017sip,Alanne:2018brf} for the case of one Higgs doublet, while the $t$-channel, although present in standard type-I seesaw leptogenesis, is usually neglected since it does not influence the washout term. In fact, none of these scattering contributions enter the expression for $W(z)$. In this work we present the general and complete scattering formulas for the aforementioned processes (see appendix~\ref{sec:scatterings}) and consider them in our numerical analysis of section~\ref{sec:BAUmodel}. The $S_{ij}$ coefficients accounting for the $\Delta N = 2$ scatterings are then given by [see eqs.~\eqref{eq:equi} and~\eqref{eq:Scatteringgen}]
    \begin{align}
        S_{i j} &= \frac{\gamma^\text{eq}_{i j}}{n_{N_i}^\text{eq} n_{N_j}^\text{eq}H(z) z} , \nonumber \\
        \gamma^\text{eq}_{i j} &= \sum_{a,b=1}^{n_H}  \gamma^\text{eq}(N_i N_j \leftrightarrow \Phi_a \Phi_b) + \sum_{p,l=1}^{2 n_S} \gamma^\text{eq}(N_i N_j \leftrightarrow h_p h_l),
    \end{align}
which are computed using eq.~\eqref{eq:reacscattering} and the cross sections obtained in appendix~\ref{sec:scatterings}.
    
\item \textbf{Washout:}
    
The washout term only accounts for LNV processes, since these alter the ($B-L$) asymmetry, such as inverse decays $\ell_\alpha \Phi_a \rightarrow N_i$ and $\Delta L=1,2$ scatterings. Hence, the additional $N_j \rightarrow N_i h_k$ decays or $\Delta N =2$ scattering processes do not contribute to~$W(z)$. As mentioned before, we will not consider $\Delta L=2$ scatterings. Thus,
    \begin{align}
    W(z) &= W_{\rm ID}(z) + W_{\Delta L=1}(z) \; ,
    \end{align}
    with [see eq.~\eqref{eq:InvDecaygen}],
    \begin{align}
    W_{\rm ID}(z) &= \sum_{i=1}^{n_R} \frac{1}{2} \frac{N_{N_i}^{\text{eq}}}{N_{\ell}^{\text{eq}}} D_i = \frac{1}{4} \sum_{i=1}^{n_R} K_i \sqrt{r_{i 1}} z_i^3 \mathcal{K}_1(z_i)  \; , \\
    W_{\Delta L=1}(z) &= \sum_{i=1}^{n_R} \frac{N_{N_i}^{\text{eq}}}{N_{\ell}^{\text{eq}}} \left(S_i^\prime + \frac{N_{N_i}}{N_{N_i}^{\text{eq}}} S_i^{\prime \prime} \right) \; ,
    \end{align}
    where $N_{\ell}^{\text{eq}}=3/4$ [see eq.~\eqref{eq:ournotation}] and 
    \begin{align}
        S_i^\prime = \frac{\gamma_i^{\text{eq} \; \prime}}{n_{N_i}^\text{eq} H(z) z} \; , \; \gamma_i^{\text{eq} \; \prime} &= \sum_{\alpha=e, \mu, \tau} \left[ \gamma^\text{eq}(N_i t_R  \leftrightarrow \ell_{\alpha} q_L) + \gamma^\text{eq}(N_i q_L \leftrightarrow \ell_{\alpha} t_R) \right]  \\ 
        & + \sum_{\alpha=e, \mu, \tau} \sum_{a=1}^{n_H} \sum_{k=1}^{2 n_S} \left[\gamma^\text{eq}(N_i h_k \leftrightarrow \ell_{\alpha} \Phi_a) + \gamma^\text{eq}(N_i \Phi_a \leftrightarrow \ell_{\alpha} h_k) \right] \; , \nonumber \\
        S_i^{\prime \prime} = \frac{\gamma_i^{\text{eq} \; \prime \prime}}{n_{N_i}^\text{eq} H(z) z} \; , \; \gamma_i^{\text{eq} \; \prime \prime} &= \sum_{\alpha=e, \mu, \tau} \gamma^\text{eq}(N_i \ell_{\alpha} \leftrightarrow q_L t_R) \nonumber \\
        & + \sum_{\alpha=e, \mu, \tau} \sum_{a=1}^{n_H} \sum_{k=1}^{2 n_S} \gamma^\text{eq}(N_i \ell_{\alpha} \leftrightarrow \Phi_a h_k)\; ,
    \end{align}
    where $S_i^\prime$ ($S_i^{\prime \prime})$ contains all $\Delta L = 1$ scattering processes where the charged-lepton $\ell_\alpha$ appears in the final (initial) state.
    
\end{itemize}

Up to now we have set our theoretical framework and presented the expressions for the CP asymmetries and the unflavoured BEs needed for the computation of the BAU. In the next section, we will present a simple model where leptogenesis is entirely due to the new scalar interactions fed by a single CP-violating phase of that scalar VEV. It turns out that (Dirac and Majorana) low-energy CP violation in the neutrino sector is also induced.

\section{A simple model for leptogenesis with high-energy SCPV}
\label{sec:model}

\begin{table}[t!]
\renewcommand{\arraystretch}{1.1}
	\centering
	\begin{tabular}{| K{1.8cm} | K{1.5cm} | K{3.0cm} |  K{2.0cm} | K{2.0cm} | K{2.0cm} | }
		\hline 
&Fields&\EW&  $\mathcal{Z}_{8}^{e}$ &  $\mathcal{Z}_{8}^{\mu}$  &  $\mathcal{Z}_{8}^{\tau}$ \\
		\hline 
		\multirow{8}{*}{Fermions} 
&$\ell_{e L} $&($\mathbf{2}, {-1/2}$)   & {$\omega^5$} & {$\omega^7$}  & {$\omega^6$}    \\
&$\ell_{\mu L}$&($\mathbf{2}, {-1/2}$)  & {$\omega^7$} & {$\omega^5$}  & {$\omega^5$}     \\
&$\ell_{\tau L}$&($\mathbf{2}, {-1/2}$) & {$\omega^6$} & {$\omega^6$}  & {$\omega^7$}     \\
&$e_R$&($\mathbf{1}, {-1}$)     & {$\omega^4$} & {$\omega^7$}  & {$\omega^6$}     \\
&$\mu_R$&($\mathbf{1}, {-1}$)   & {$\omega^7$} & {$\omega^4$}  & {$\omega^4$}    \\
&$\tau_R$&($\mathbf{1}, {-1}$)  & {$\omega^6$} & {$\omega^6$}  & {$\omega^7$}    \\
&$\nu_{R_1}$&($\mathbf{1}, {0}$)& {$\omega^6$} & {$\omega^6$}& {$\omega^6$}    \\
&$\nu_{R_2}$&($\mathbf{1}, {0}$)& {$1$}    & {$1$}    & {$1$}   \\
		\hline 
		\multirow{3}{*}{Scalars}
&$\Phi_1$&($\mathbf{2}, {1/2}$)&\multicolumn{3}{c|}{{$1$}}\\
&$\Phi_2$&($\mathbf{2}, {1/2}$)&\multicolumn{3}{c|}{{$\omega$}}\\
&$S$&($\mathbf{1}, {0}$)&\multicolumn{3}{c|}{{\ $\omega^2$}}\\
\hline
	\end{tabular}
	\caption{Field content of the model and corresponding transformation properties under the \EW~gauge group. For the $\mathcal{Z}_8$ symmetry we have $\omega^k = e^{ik\pi/4}$.}
	\label{tab:part&sym} 
\end{table}

We now focus on a simple realisation of the general framework described in the previous sections, i.e. a SM extension with two RH neutrinos $\nu_{R_{1,2}}$ ($n_R=2$) and one complex scalar singlet~$S$ ($n_S=1$). With this setup, SCPV can be achieved at high energies through the complex VEV of the scalar singlet $S$, provided we guarantee that the $S^4$ term in the scalar potential is present~\cite{Branco:1999fs}. As also noted in refs.~\cite{Correia:2019vbn,Camara:2020efq}, at least two-Higgs doublets are necessary to implement Abelian flavour symmetries in this type of scenarios. Thus, we will add an extra scalar doublet to the field content of our model such that $n_H=2$. With $n_R=2$, $n_S=1$ and $n_H=2$, the most restrictive Abelian symmetry which can be implemented is a~$\mathcal{Z}_{8}$ which, as we will see shortly, will lead to testable low-energy predictions. 

The $\mathcal{Z}_{8}$ symmetry is suitable for two reasons: i) it allows for the $S^4$ term in the scalar potential needed for SCPV, provided $S$ transforms as $S \rightarrow \omega^2 S$ [$\omega=\text{exp}(i \pi / 4)$] and ii) with the aforementioned minimal particle content the $\mathcal{Z}_8$ is the lowest-order $\mathcal{Z}_n$ symmetry containing a sufficient number of charges to obtain non-trivial flavour predictions. The particle content of the model, together with the $\mathcal{Z}_{8}$ charge assignments, is summarised in table~\ref{tab:part&sym}. The three cases $\mathcal{Z}_{8}^{e}$, $\mathcal{Z}_{8}^{\mu}$ and $\mathcal{Z}_{8}^{\tau}$ differ from each other by the $\mathcal{Z}_{8}$ charges of the SM lepton fields. As will be clear in the next sections, this minimal model illustrates the main idea of this work: the single VEV phase $\theta$ of $S$ provides a common source for CP violation required to generate the BAU and low-energy leptonic CP violation.

\subsection{Compatibility with low-energy neutrino data}
\label{sec:neutrinodata}

With $n_{R,H}=2$ and $n_S=1$, the relevant couplings and mass parameters of eq.~\eqref{eq:LYuk} are $\Yl^{1,2}$, $\YD^{1,2}$, $\YR^\prime$ and $\MR^0$ which are all real, since we impose CP symmetry at the Lagrangian level. As already mentioned, the three $\mathcal{Z}_8$ lepton charge assignments given in table~\ref{tab:part&sym} will lead to different Yukawa couplings $\Yl^{1,2}$ and $\YD^{1,2}$. For the specific $\mathcal{Z}_8^{\mu}$ case, we have
\begin{align}
    \Yl^1=\begin{pmatrix}
    y_1&0&0\\
    0&0&0\\
    0&0&y_4
    \end{pmatrix},\; 
    &\Yl^2=\begin{pmatrix}
    0&0&y_2\\
    0&y_3&0\\
    0&0&0
    \end{pmatrix},\;
    \YD^1=\begin{pmatrix}
    0 & 0\\
    0 & 0\\
    y_{D_3} & 0
    \end{pmatrix},\;
   \YD^2=\begin{pmatrix}
    0 & y_{D_1}\\
    y_{D_2} & 0 \\
    0 & 0
    \end{pmatrix}, \nonumber \\[0.2cm]
    &\MR^0=\begin{pmatrix}
    0&0\\
    .&m_R
    \end{pmatrix},\; 
    \YR'=\begin{pmatrix}
    0 & y_{R_S}\\
    .&0
    \end{pmatrix},\;
    \label{eq:yukawastructures}
\end{align}
where all parameters are real and the dots reflect the symmetric nature of the Majorana matrices. The corresponding matrices for $\mathcal{Z}_8^{e}$ and $\mathcal{Z}_8^{\tau}$ are obtained from the above by performing the permutations $\ell_L,e_R \rightarrow \mathbf{P}_{12}\, \ell_L,e_R$ and $\ell_L,e_R \rightarrow \mathbf{P}_{23}\, \ell_L,e_R$, respectively, being
\begin{equation}
\mathbf{P}_{12} = \begin{pmatrix} 
0 & 1 & 0 \\ 
1 & 0 & 0 \\
0 & 0 & 1
\end{pmatrix} \; , \; \mathbf{P}_{23} = \begin{pmatrix} 
1 & 0 & 0 \\ 
0 & 0 & 1 \\
0 & 1 & 0 
\end{pmatrix} \; .
\label{eq:P12P23}
\end{equation}
Notice that, due to the~$\mathcal{Z}_8$ symmetry, the term $\overline{\nu_R} \nu_R^c S$ is absent from the Lagrangian. However, since the $\overline{\nu_R} \nu_R^cS^\ast$ coupling $y_{R_S}$ is $\mathcal{Z}_{8}$ invariant, LCPV can, in principle, be successfully transmitted to the neutrino sector as long as~$S$ acquires a complex VEV. In fact, the charge assignments guarantee that $\MR^0$ and $\YR^\prime$ are not proportional to each other implying that $\Ur$ will be complex. Consequently, as we will see shortly, LCPV probed in neutrino oscillation experiments originates dynamically from the vacuum and so does the BAU. 

We consider the following VEV assignments for the neutral components of the two scalar doublets $\phi^0_{1,2}$ and for the complex singlet $S$:
\begin{align}
    \langle \phi^0_{1} \rangle = \frac{v_1}{\sqrt{2}} \;,\;
    \langle \phi^0_{2} \rangle = \frac{v_2}{\sqrt{2}} \;,\; 
    \langle S \rangle=\frac{u e^{i\theta}}{\sqrt{2}} \;,
    \label{eq:vevs}
\end{align}
where $v_{1,2}$, $u$ and $\theta$ are real. The charged-lepton, Dirac neutrino and RH neutrino mass matrices, needed to compute low-energy neutrino masses and mixing, are given by:
\begin{align}
   \Ml =\begin{pmatrix}
    a_1&0&a_2\\
    0&a_3&0\\
    0&0&a_4
    \end{pmatrix},\;
   \MD=\begin{pmatrix}
    0 & m_{D_1}\\
    m_{D_2} & 0 \\
    m_{D_3} & 0
    \end{pmatrix}, \;
    \MR=\begin{pmatrix}
    0& m_{R_S} e^{-i \theta}\\
    .&m_R
    \end{pmatrix} \; ,
    \label{eq:masstructures}
\end{align}
where
\begin{equation}
\begin{aligned}
a_{1,4} = \frac{v_1 y_{1,4}}{\sqrt{2}} \; , \; a_{2,3} = \frac{v_2 y_{2,3}}{\sqrt{2}} \; , \; m_{D_{1,2}} = \frac{v_2 y_{D_{1,2}}}{\sqrt{2}} \; , \; m_{D_{3}} = \frac{v_1 y_{D_{3}}}{\sqrt{2}} \; , \; m_{R_S} = \frac{u y_{R_S}}{\sqrt{2}} \,.
\label{eq:pdef}
\end{aligned}
\end{equation}
The specific form of $\Ml$ indicates that $a_3$ is directly the mass of one physical charged lepton, which we call $\ell_2$, and is decoupled from the other two $\ell_{1,3}$. Three out of the four $a_i$ can be written in terms of the charged-lepton masses $m_{\ell_{1,2,3}}$ and a single $a$ parameter, considered here to be $a_4$. Thus, we have
\begin{equation}
a_1^2 = \frac{m^2_{\ell_1} m^2_{\ell_3}}{a_4^2}, \  a_2^2 = \frac{(a_4^2-m^2_{\ell_1}) (m^2_{\ell_3}-a_4^2)}{a_4^2}, \ a_3^2 = m^2_{\ell_2}, \ m_{\ell_1}^2<a_4^2<m_{\ell_3}^2,
\end{equation}
The unitary matrix $\mathbf{V}_{L}$ that diagonalises the Hermitian matrix $\mathbf{H}_{\ell} = \mathbf{M}_{\ell} \mathbf{M}_{\ell}^{\dagger}$ is given by\footnote{Since we imposed CP at the Lagrangian level, $\mathbf{V}_{L}$ and $\mathbf{H}_{\ell}$ are actually orthogonal and symmetric matrices, respectively.}
\begin{align}
\mathbf{H}_{\ell} = \begin{pmatrix} 
a_1^2 + a_2^2 & 0 & a_2 a_4\\ 
0 & a_3^2  & 0 \\
a_2 a_4 & 0 & a_4^2    
\end{pmatrix}\;,\; \mathbf{V}_{L} = \begin{pmatrix} 
c_L & 0 & s_L\\ 
0 & 1 & 0 \\
- s_L & 0 & c_L
\end{pmatrix} \,,
\label{eq:VLmodel}
\end{align}
with $c_{L}\equiv \cos \theta_{L}$, $s_{L}\equiv \sin \theta_{L}$ and
\begin{equation}
\tan \left(2 \theta_{L}\right) = \dfrac{2 \sqrt{(a_4^2-m_{\ell_1}^2)(m_{\ell_3}^2-a_4^2)}}{(m_{\ell_1}^2+m_{\ell_3}^2)-2 a_4^2} \,.
\label{eq:tLexp}
\end{equation}
For each symmetry charge assignment $\mathcal{Z}_8^{e, \mu, \tau}$, one must consider the three possible choices for $\ell_2$, i.e. $\ell_2= e, \mu, \tau$. For the $\mathcal{Z}_8^{\mu}$ case, $\mathbf{V}_L$ in eq.~\eqref{eq:VLmodel}  corresponds to $\ell_2 = \mu$. For electron (tau) decoupled i.e., $\ell_2 = e$ ($\ell_2 = \tau$), $\mathbf{V}_L$ is replaced by $\mathbf{P}_{12}\mathbf{V}_L$ ($\mathbf{P}_{23}\mathbf{V}_L$), with $\mathbf{P}_{12}$ $(\mathbf{P}_{23})$ given in eq.~\eqref{eq:P12P23} and $\theta_L$ determined by eq.~\eqref{eq:tLexp}.

Taking into account eqs.~\eqref{eq:typeI} and \eqref{eq:masstructures}, the effective neutrino mass matrix in the flavour basis reads
\begin{align}
\Mnu = 
e^{2 i \theta}\begin{pmatrix} 
0 & y\, e^{-i \theta}  & y \sqrt{\dfrac{z}{x}}\, e^{-i \theta}  \\
. &  x &  \sqrt{x z} \\
. & . &  z
\end{pmatrix}\;,\;
x=\frac{m_R m_{D_2}^2}{m_{R_S}^2}\;,\;
y=-\frac{m_{D_1}m_{D_2}}{m_{R_S}}\;,\;
z=\frac{m_R m_{D_3}^2}{m_{R_S}^2}\;.
\label{eq:Mnuxyz}
\end{align}
Performing the rotation to the charged-lepton mass basis with the unitary matrix $\mathbf{V}_{L}$ given in eq.~\eqref{eq:VLmodel}, we obtain for $\mathcal{Z}_8^{\mu}$:
\begin{align}
&\V_L^T \Mnu \V_L=  \label{eq:mnuclmassbasis}\\
 & \begin{pmatrix} 
 - e^{-i \theta} y \sqrt{\dfrac{z}{x}} \sin(2 \theta_L) + z s_L^2\quad\quad & e^{-i \theta} y c_L-\sqrt{x z} s_L \quad\quad& e^{-i \theta} y \sqrt{\dfrac{z}{x}} \cos(2 \theta_L) - \dfrac{z}{2} \sin(2 \theta_L) \\
. & x &  e^{-i \theta} y s_L + \sqrt{x z} c_L\\
. & . & e^{-i \theta} y \sqrt{\dfrac{z}{x}} \sin(2 \theta_L) + z c_L^2
\end{pmatrix} \; , \nonumber
\end{align}
while for $\mathcal{Z}_8^e$ ($\mathcal{Z}_8^\tau$) permutation  $\mathbf{P}_{12}$ ($\mathbf{P}_{23}$) in eq.~\eqref{eq:P12P23} must be applied both on the left and right. We remark that the $\mathcal{Z}_8$ flavour symmetry reduces the total number of free effective parameters to five. Namely, $\theta_L$ (or equivalently $a_4$) from $\Ml$ and $(x,y,z,\theta)$ from $\Mnu$. Notice that the singlet VEV phase $\theta$ cannot be removed via field redefinitions and, consequently, it may successfully lead to low-energy CP violation in the neutrino sector. Furthermore, since there are only two RH neutrinos, one of the light neutrinos is massless. This is readily seen by computing the eigenvalues of the matrices in eqs.~\eqref{eq:Mnuxyz} and~\eqref{eq:mnuclmassbasis}.
\begin{table}[!t]
\renewcommand{\arraystretch}{1.2}
\centering
\setlength{\tabcolsep}{10pt}
\begin{tabular}{|l|c|c|}  
\hline
Parameter  & Best Fit $\pm 1 \sigma$ & $3\sigma$ range \\ \hline
$\theta_{12} (^\circ)$ & $34.3\pm1.0$ &  $31.4 \rightarrow 37.4$ \\
$\theta_{23} (^\circ) [\text{NO}]$ & $49.26\pm0.79$ &  $ 41.20 \rightarrow 51.33 $ \\
$\theta_{23} (^\circ) [\text{IO}]$ & $49.46^{+0.60}_{-0.97}$  &  $ 41.16 \rightarrow 51.25$ \\
$\theta_{13} (^\circ) [\text{NO}]$ & $8.53^{+0.13}_{-0.12}$ &  $8.13 \rightarrow 8.92$\\
$\theta_{13} (^\circ) [\text{IO}]$ & $8.58^{+0.12}_{-0.14}$ &  $8.17 \rightarrow 8.96$ \\
$\delta  (^\circ) [\text{NO}]$ & $194^{+24}_{-22}$ & $128 \rightarrow 359 $ \\
$\delta  (^\circ) [\text{IO}]$ & $284^{+26}_{-28}$ &  $200 \rightarrow 353 $ \\
$\Delta m_{21}^2 \left(\times 10^{-5} \ \text{eV}^2\right)$ & $7.50^{+0.22}_{-0.20}$ &  $ 6.94 \rightarrow 8.14$ \\
$\left|\Delta m_{31}^2\right| \left(\times 10^{-3}  \ \text{eV}^2\right) [\text{NO}]$ & $2.55^{+0.02}_{-0.03}$  & $2.47 \rightarrow 2.63 $ \\
$\left|\Delta m_{31}^2\right| \left(\times 10^{-3} \ \text{eV}^2\right) [\text{IO}]$ & $2.45^{+0.02}_{-0.03}$ & $2.37 \rightarrow 2.53$\\
\hline
\end{tabular}
\caption{Current allowed intervals for the lepton mixing angles, neutrino mass-squared differences and the Dirac phase $\delta$ obtained from the global fit of neutrino oscillation data performed in ref.~\cite{deSalas:2020pgw} (see also refs.~\cite{Esteban:2020cvm} and~\cite{Capozzi:2021fjo}).}
\label{tab:dataref}
\end{table}

In order to test the compatibility of our model with neutrino oscillation data, the mass matrix $\Mnu$ of eq.~\eqref{eq:mnuclmassbasis} for the $\mathcal{Z}_8^{e,\mu,\tau}$ cases must be matched with that defined through low-energy parameters, namely
\begin{align}
    \Mnuh=\U^\ast\text{diag}(m_1,m_2,m_3)\U^\dagger,
\end{align}
where $\U$ is the matrix in eq.~\eqref{eq:leptonmix}. For massive Majorana neutrinos, $ \mathbf{U}$ can be parameterised by three mixing angles $\theta_{12}$, $\theta_{23}$, and $\theta_{13}$, and three CP-violating phases: a Dirac-type phase $\delta$ and two Majorana-type phases $\alpha_{21}$ and $\alpha_{31}$:
\begin{align}
\U=\begin{pmatrix}
c_{12}c_{13}&s_{12}c_{13}&s_{13}\\
-s_{12}c_{23}-c_{12}s_{23}s_{13}e^{i\delta}&c_{12}c_{23}-s_{12}s_{23}s_{13}e^{i\delta}&s_{23}c_{13}e^{i\delta}\\
s_{12}s_{23}-c_{12}c_{23}s_{13}e^{i\delta}&-c_{12}s_{23}-s_{12}c_{23}s_{13}e^{i\delta}&c_{23}c_{13}e^{i\delta}
\end{pmatrix} 
\begin{pmatrix}
1&0&0\\
0&e^{i\frac{\alpha_{21}}{2}}&0\\
0&0&e^{i\frac{\alpha_{31}}{2}}
\end{pmatrix} \; ,
\label{eq:Uparam}
\end{align}
where $c_{ij} \equiv \cos\theta_{ij}$ and $s_{ij} \equiv \sin\theta_{ij}$. Several neutrino oscillation experiments have been constraining neutrino mass and mixing parameters, namely $\Delta m_{21}^2= m_2^2 - m_1^2$ , $\Delta m_{31}^2= m_3^2 - m_1^2$, $\theta_{12}$, $\theta_{23}$, $\theta_{13}$ and $\delta$. In table~\ref{tab:dataref}, we show the results obtained from the most recent global fit of neutrino oscillation parameters~\cite{deSalas:2020pgw} (see also refs.~\cite{Esteban:2020cvm} and~\cite{Capozzi:2021fjo}). Both mass orderings are considered: normal ordering~(NO) where $m_1< m_2< m_3$, and inverted ordering~(IO) where  $m_3 < m_1< m_2$. In our case, since the lightest neutrino is massless ($m_1=0$ for NO, and $m_3=0$ for IO), the Majorana phase $\alpha_{31}$ can be rephased away, and only $\alpha\equiv\alpha_{21}-\alpha_{31}$ is physically relevant.

We tested the viability of cases $\mathcal{Z}_8^{e,\mu,\tau}$, for both NO and IO, using a standard $\chi^2$-analysis, by minimising the function
\begin{align}
    \chi^2(x,y,z,\theta,\theta_L)=\sum_i\dfrac{[\mathcal{P}_i(x,y,z,\theta,\theta_L)-\mathcal{O}_i]^2}{\sigma_i^2},
    \label{eq:chisq}
\end{align}
with respect to the neutrino observables $\Delta m^2_{ij}$, $\theta_{ij}$ and $\delta$. In the above, $\mathcal{P}_i$ corresponds to the predicted value for the observable $i$ obtained by varying the input parameters $x$, $y$, $z$, $\theta$ and $\theta_L$, while $\mathcal{O}_i$ ($\sigma_i$) denotes the correspondent best-fit value ($1\sigma$ experimental uncertainty), indicated in table~\ref{tab:dataref}.

\begin{table}[!t]
\renewcommand{\arraystretch}{1.1}
\centering
\setlength{\tabcolsep}{10pt}
\begin{tabular}{|c|c|c|c|c|c|c|c|c|}  
\hline
\multirow{2}{*}{Case}  & $\theta_{12}$ & $\theta_{13}$ & $\theta_{23}$ & \multirow{2}{*}{$\delta/\pi$} & \multirow{2}{*}{$\alpha/\pi$} & $m_{\beta\beta}$&  $m_\beta$  & $\sum_i m_i$\\
&$(^\circ)$&$(^\circ)$&$(^\circ)$&&&(meV)&(meV)&(meV)\\
\hline
$\mathcal{Z}_8^\mu$  (IO) &$35.48$&8.60&$49.62$&$1.88$&$0.92$&$16.6$&$49.2$&$99.7$\\
\hline
\end{tabular}
\caption{$\theta_{12}$, $\theta_{13}$, $\theta_{23}$, $\delta$, $\alpha$, $m_{\beta\beta}$, $m_{\beta}$ and $\sum_i m_i$ best-fit values for the only case compatible with neutrino oscillation data at the $1\sigma$ level: $\mathcal{Z}_8^\mu$ muon-decoupled with IO neutrino masses. All the remaining cases, $\mathcal{Z}_8^e$, $\mathcal{Z}_8^\tau$ and $\mathcal{Z}_8^{\mu}$ for NO, and considering all possible decoupled charged-lepton states, are not compatible with data.}
\label{tab:compatibility}
\end{table}
\begin{table}[!t]
\renewcommand{\arraystretch}{1.1}
\centering
\setlength{\tabcolsep}{10pt}
\begin{tabular}{|c|c|c|c|}  
\hline
Case  & $\theta/\pi$ & $\theta_L/\pi$&$(x,y,z)$ (meV)\\
\hline
$\mathcal{Z}_8^\mu$  (IO) &$1.89$&$7.29\times10^{-2}$&$(0.325,32.8,0.426)$\\
\hline
\end{tabular}
\caption{Values of the $\theta$, $\theta_L$, $x$, $y$ and $z$ input parameters in eq.~\eqref{eq:Mnuxyz} which lead to the best-fit case shown in table~\ref{tab:compatibility}.}
\label{tab:compatibilityoutput}
\end{table}

In table~\ref{tab:compatibility}, we show the $\theta_{12}$, $\theta_{13}$, $\theta_{23}$ and $\delta$ values for the case that best fits the data, i.e $\mathcal{Z}_8^\mu$ with muon decoupled and IO (the $\Delta m_{21,31}$ are on their best-fit values). We also indicate the predictions for the Majorana phase $\alpha$, effective masses $m_{\beta\beta}$ (neutrinoless double beta decay) and $m_\beta$ ($\beta$-decay), as well as the sum of neutrino masses~$\sum_i m_i$. The remaining cases, $\mathcal{Z}_8^\mu$ for NO and $\mathcal{Z}_8^{e,\tau}$, are compatible with data at more than $3\sigma$.  From these results, one can see that the model prefers $\theta_{23}$ in the second octant, and that $m_{\beta\beta}$, $m_\beta$ and $\sum_i m_i$ are well below the current most stringent limits from KamLAND-Zen -- $m_{\beta\beta}< (36-156)$~meV ($90\%$ CL)~\cite{KamLAND-Zen:2022tow}, KATRIN -- $m_{\beta}<0.8$~eV ($90\%$ CL)~\cite{KATRIN:2021uub}, and Planck -- $\sum_i m_i<(0.12-0.54)$~eV ($95\%$ CL)~\cite{Planck:2018vyg}, respectively. The corresponding values of the input parameters $(\theta,\theta_L,x,y,z)$ in eq.~\eqref{eq:mnuclmassbasis} are given in table~\ref{tab:compatibilityoutput} and will be used in the leptogenesis analysis of section~\ref{sec:BAUmodel}. Note that the VEV phase $\theta$ is close to $340^\circ$ and the Dirac CP phase is predicted to be $\delta \simeq \theta$, which is $1.5\sigma$ away from the experimental best-fit for $\delta$. This relation between $\theta$ and $\delta$ stems from the underlying $\mathcal{Z}_8$ symmetry of our model. It is indeed remarkable that, with such a simple setup, a one-to-one correspondence between $\delta$ (low-energy Dirac CP phase) and $\theta$ (vacuum CP phase) can be established, connecting two completely different sectors of the model.

\subsection{BAU generation from vacuum CP violation}
\label{sec:BAUmodel}

In the model under consideration, CP is spontaneously broken at high energies by the complex VEV of the singlet $S$. The $\mathcal{Z}_8$-invariant scalar potential is\footnote{We added the soft-breaking term $\propto (\Phi_1^\dagger \Phi_2)$ to avoid a massless Goldstone boson which would stem from the Higgs doublets neutral degrees of freedom after EWSB.}
\begin{align}
V(\Phi_1,\Phi_2,S) &= m^2_{1} (\Phi_1^\dagger \Phi_1 ) + m^2_{2} (\Phi_2^\dagger \Phi_2 ) + m^2_{1 2} \left[ (\Phi_1^\dagger \Phi_2 ) + (\Phi_2^\dagger \Phi_1 ) \right] \nonumber \\
&+ \frac{\lambda_{1}}{2} (\Phi_1^\dagger \Phi_1 )^2 + \frac{\lambda_{2}}{2} (\Phi_2^\dagger \Phi_2 )^2 + \lambda_{3} (\Phi_1^\dagger \Phi_1 ) (\Phi_2^\dagger \Phi_2 ) + \lambda_{4} (\Phi_1^\dagger \Phi_2 ) (\Phi_2^\dagger \Phi_1) \nonumber \\
& + \lambda_{1 S} (\Phi_1^\dagger \Phi_1 ) |S|^2 + \lambda_{2 S} (\Phi_2^\dagger \Phi_2 ) |S|^2 \nonumber \\
&+ m^2_{S} |S|^2 + m^{\prime \; 2}_{S} \left(S^2 + {S^\ast}^2\right)  + \frac{\lambda_S}{2} |S|^4 + \lambda_S^\prime \left(S^4 + {S^\ast}^4 \right) \; ,
\label{eq:potentialmodel}
\end{align}
where all parameters are real, since CP invariance is imposed. We consider that at high-energies or, in other words, at temperatures much higher than the EW scale, only $S$ has non-zero VEV. In fact, within the unflavoured leptogenesis scenario considered in this work, the heavy Majorana neutrino masses are such that $M_{1,2}\sim u \gtrsim 10^{12}$~GeV. Hence, the scalar $S$ is naturally decoupled from $\Phi_{1,2}$. For the above potential, we then obtain three non-trivial scalar potential minimisation conditions:
\begin{align}
\text{(i) :} & \;  m_{S}^2 = - \frac{1}{2} \left[u^2 \left(\lambda_{S} + 4 \lambda_{S}^{\prime}\right) + 4 m_{S}^{\prime} \right] \; , \; \theta = k \pi \; , k \in \mathbb{Z} \; ; \\
\text{(ii) :} & \; m_{S}^2 = - \frac{1}{2} \left[u^2 \left(\lambda_{S} + 4 \lambda_{S}^{\prime}\right) - 4 m_{S}^{\prime} \right] \; , \; \theta = \frac{\pi}{2} + k \pi \; , k \in \mathbb{Z} \; ; \\
\text{(iii) :} & \; m_{S}^2 = - \frac{u^2}{2} \left(\lambda_{S} - 4 \lambda_{S}^{\prime} \right) \; , \; \cos (2 \theta) = - \frac{m^{\prime\,2}_S}{2 u^2 \lambda_{S}^{\prime}} \; .
\label{eq:SCPVsol}
\end{align}
Notice that, in spite of (ii) leading to $\theta = \pi/2$, it can be shown that in this case the vacuum does not violate CP~\cite{Branco:1999fs}. Therefore, the only viable solution to implement SCPV is (iii). In the exact $\mathcal{Z}_8$-symmetric limit, i.e. if the soft-breaking parameter $m^{\prime\,2}_S$ vanishes, we have $\theta= \pi/4+k\pi/2$, still leading to SCPV. However, such value of $\theta$ is not compatible with neutrino data\footnote{For this case, the solution which fits the data the best corresponds to $\theta=7\pi/4$, leading to best-fit value(s) of  $\theta_{12}$ ($\theta_{13}$ and $\theta_{23}$) which is (are) approximately at $3\sigma$ ($1\sigma$) distance of its (their) experimental best fit.} (see table~\ref{tab:compatibilityoutput}). Therefore, a non-vanishing $m^{\prime\,2}_S$ is required, which also helps avoiding the domain wall problem arising from the spontaneous breaking of the $\mathcal{Z}_8$ discrete symmetry. The SCPV solution (\ref{eq:SCPVsol}) corresponds to the global minimum of the potential if $ \left(m_{S}^{\prime \; 4} - 4 u^4 \lambda_{S}^{\prime \; 2} \right) / (4 \lambda_{S}^{\prime}) > 0$. Also, boundedness from below requires
\begin{align}
&\lambda_1, \lambda_2, \lambda_S > 0 \; , \nonumber \\
& \lambda_3 + \sqrt{\lambda_1 \lambda_2} > 0 \; , \; \lambda_3 + \lambda_4 + \sqrt{\lambda_1 \lambda_2} > 0 \; , \; \lambda_S - 4 |\lambda_S^\prime| > 0 \; , \nonumber \\
& \lambda_{1 S} + \sqrt{\lambda_1 \lambda_S} > 0 \; , \; \lambda_{2 S} + \sqrt{\lambda_2 \lambda_S} > 0 \; , \nonumber \\ &\lambda_{1 S} + \sqrt{\lambda_1 \left(\lambda_S - 4 |\lambda_S^\prime|\right)} > 0 \; , \; \lambda_{2 S} + \sqrt{\lambda_2 \left(\lambda_S - 4 |\lambda_S^\prime|\right)} > 0 \; .
\label{eq:bfb}
\end{align}

In section~\ref{sec:leptogenesis}, we have computed the leptogenesis CP asymmetries in the fermion and $S$ scalar mass-eigenstate basis. For the model under discussion with $n_S=1$, $\bm{\mathcal{M}}_S^2$ and $\mathbf{V}$ in eqs.~\eqref{eq:mixS} and~\eqref{eq:mixingscalar} are $2\times2$ matrices. Namely, in the $(S_R,S_I)$ basis we have,
\begin{align}
\bm{\mathcal{M}}_S^2 = u^2 \begin{pmatrix}
(\lambda_S + 4 \lambda_S^\prime ) \cos^2 \theta & (\lambda_S - 12 \lambda_S^\prime ) \cos \theta \sin \theta\\
(\lambda_S - 12 \lambda_S^\prime ) \cos \theta \sin \theta & (\lambda_S + 4 \lambda_S^\prime ) \sin^2 \theta
\end{pmatrix} \; ,
\end{align}
leading to the $h_{1,2}$ scalar masses
\begin{align}
m_{h_{1,2}}^2 = \frac{u^2}{2} \left(\lambda_S + 4 \lambda_S^\prime \mp \sqrt{\lambda_S^2 - 8 \lambda_S \lambda_S^\prime + 80 \lambda_S^{\prime 2} + 16 (\lambda_S - 4 \lambda_S^\prime ) \lambda_S^\prime \cos( 4 \theta) } \right)\,.
\label{eq:massesh1h2}
\end{align}

These states will be responsible for new contributions to the CP asymmetry, as seen in section \ref{sec:asymmetry}. In turn, scalar mixing is encoded by
\begin{align}
\mathbf{V} = \begin{pmatrix}
\cos \theta_S & \sin \theta_S\\
 -\sin \theta_S & \cos \theta_S
\end{pmatrix} \; , \; \tan(2 \theta_S) = - \frac{(\lambda_S - 12 \lambda_S^\prime) \tan (2 \theta)}{\lambda_S + 4 \lambda_S^\prime} \; .
\label{eq:thetaS1}
\end{align}
Inverting eq.~\eqref{eq:massesh1h2}, one can write $\lambda_S$ and $\lambda_S'$ in terms of $m_{h_{1,2}}$ to find
\begin{align}
\tan(2 \theta_S) &= \mp \frac{\sqrt{1+r_h^4 - 6 r_h^2 - \left(1+r_h^2\right)^2 \cos(4 \theta)}}{\sqrt{2} \left(1 + r_h^2\right)\cos(2 \theta)} \;,\;r_h\equiv \dfrac{m_{h_1}}{m_{h_2}}\,,
\label{eq:thetaS2}
\end{align}
where the minus (plus) sign in $\mp$ is valid when $\sin(2\theta)$ is positive (negative). By requiring $\theta_S$ to be real, we get the following condition for $r_h$:
\begin{align}
r_h<{\rm min}\left\{|\tan\theta|,|\cot\theta|\right\}\,,
\label{eq:mh1mh2ratiolimit}
\end{align}
which, using the best-fit value of $\theta$ indicated in table~\ref{tab:compatibilityoutput} for $\mathcal{Z}_8^\mu$, leads to $m_{h_1}/m_{h_2}\lesssim 0.37$.
Once more, we see that requiring compatibility with neutrino data imposes constraints on the scalar sector of the model, which is an interesting and uncommon feature.

The rotation to the heavy-neutrino mass-eigenstate basis is obtained by diagonalising $\M_R$ in eq.~\eqref{eq:masstructures} with the unitary matrix $\Ur$ defined as
\begin{align}
\Ur = \begin{pmatrix}
e^{-i \theta} \cos \theta_R & e^{-i \theta} \sin \theta_R\\
 -\sin \theta_R & \cos \theta_R
\end{pmatrix} \; , \; \tan(2 \theta_R) = 2 \frac{\sqrt{M_1 M_2}}{M_2-M_1} \; ,
\label{eq:mixMR}
\end{align}
being the masses of the heavy Majorana neutrinos $N_{1,2}$ given by:
\begin{align}
M_{1,2}^2 = \frac{1}{2} \left(m_R^2 + u^2 y_{R_S}^2 \mp m_R \sqrt{m_R^2 + 2 u^2 y_{R_S}^2} \right).
\end{align}

\begin{figure}[t!]
    \centering
   \hspace*{-1.5cm}\includegraphics[scale=0.52]{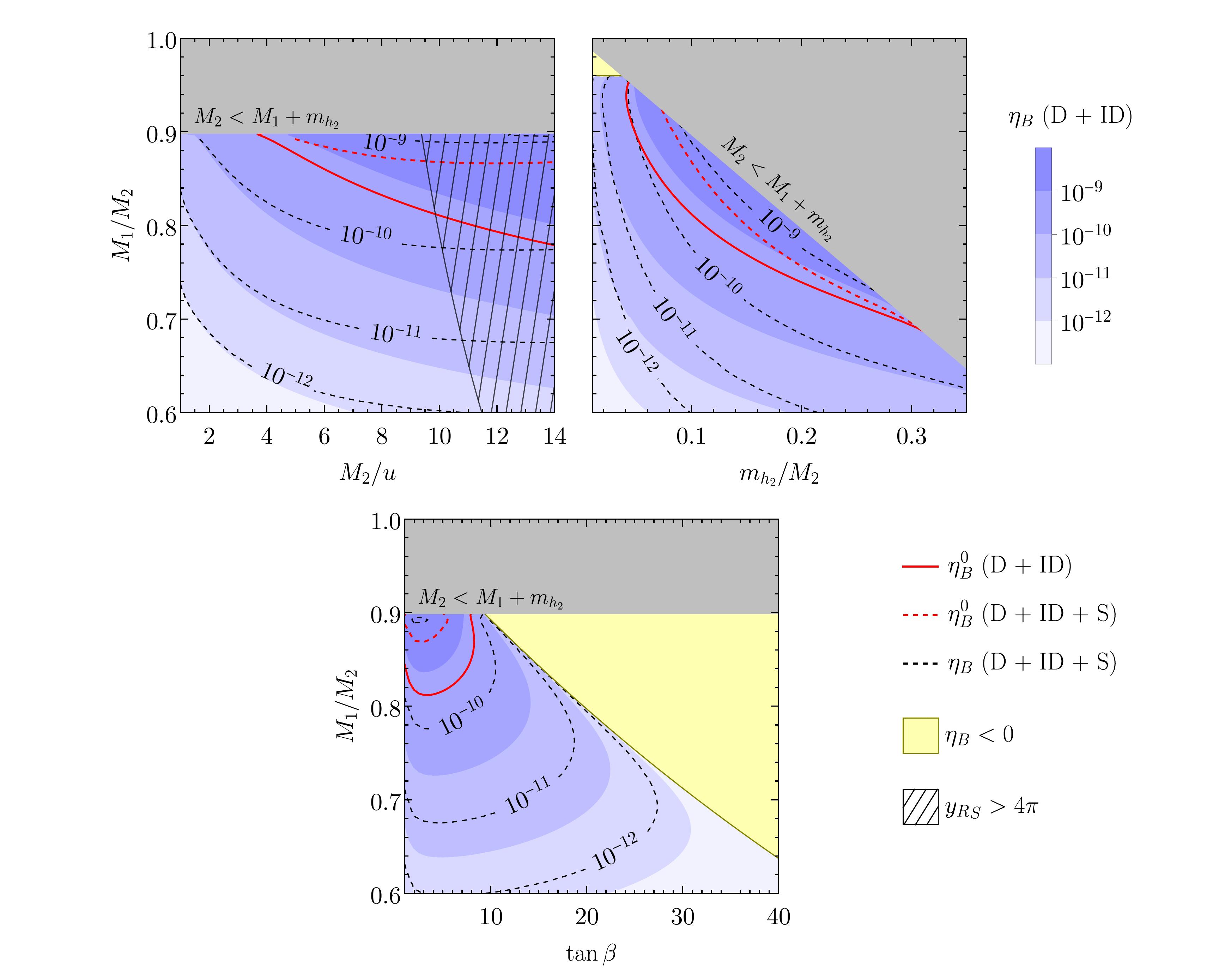}
    \caption{Top left: $\eta_B$ in the plane ($M_2/u,M_1/M_2$) with $m_{h_2}/M_2=0.1$ and $t_\beta=3$. Top right: $\eta_B$ in the plane ($m_{h_2}/M_2,M_1/M_2$) with $y_{R_S}=4\pi$ and $t_\beta=3$. Bottom: $\eta_B$ in the plane ($t_\beta,M_1/M_2$) with $y_{R_S}=4\pi$ and $m_{h_2}/M_2=0.1$. In all plots, $x$, $y$, $z$, $\theta_L$ and $\theta$ are fixed to their best-fit value for case $\mathcal{Z}_8^\mu$ (see table~\ref{tab:compatibilityoutput}), $m_{h_1}/m_{h_2}=0.37$ and $u=10^{12}$~GeV. In the grey regions one has $M_2<M_1+m_{h_2}$, while in the hatched area $y_{R_S}>4\pi$. In the yellow region $\eta_\text{B}<0$. The black dashed contours and the blue regions correspond to $\eta_B$ values computed with and without $\Delta L=1$ and $\Delta N=2$ scattering terms in the BEs, respectively. The red dashed (solid) line indicates the observed baryon asymmetry [see eq.~\eqref{eq:etab0}] when the same terms are (not) considered.}
    \label{fig:etaBwDeltaL=1}
\end{figure}

Finally, in the charged-lepton and heavy-neutrino  mass basis, one can use eqs.~\eqref{eq:YHdef}, \eqref{eq:yukawastructures}, \eqref{eq:pdef}, \eqref{eq:VLmodel}, \eqref{eq:Mnuxyz}, and \eqref{eq:mixMR} to write the two Dirac neutrino Yukawa coupling matrices  $\Y^{1,2}$ as
\begin{gather}
    \Y^1=\dfrac{\sqrt{2zM_1}}{v_1\sqrt{1-r_{12}}}\begin{pmatrix}
        s_Le^{i\theta}\quad\quad&\sqrt[4]{r_{12}}s_Le^{i\theta}\\[0.5cm]
        0\quad\quad&0\\[0.5cm]
        -c_Le^{i\theta}\quad\quad&-\sqrt[4]{r_{12}}c_Le^{i\theta}
    \end{pmatrix}, \label{eq:Y1}\\[0.5cm]
    \Y^2=\dfrac{\sqrt{2xM_1}}{v_2\sqrt{1-r_{12}}}\begin{pmatrix}
        -\dfrac{y}{x}\left(1-\sqrt{r_{12}}\right)c_L\quad\quad&\dfrac{y}{x}\dfrac{1-\sqrt{r_{12}}}{\sqrt[4]{r_{12}}}c_L\\[0.5cm]
        - e^{i\theta}\quad\quad&-\sqrt[4]{r_{12}}e^{i\theta} \\[0.5cm]
      -\dfrac{y}{x}\left(1-\sqrt{r_{12}}\right)s_L\quad\quad&\dfrac{y}{x}\dfrac{1-\sqrt{r_{12}}}{\sqrt[4]{r_{12}}}s_L
    \end{pmatrix} \; , \; r_{12}=M_1^2/M_2^2\;,
    \label{eq:Y2}
\end{gather}
for case $\mathcal{Z}_8^\mu$. Furthermore, since $\YR=\mathbb{0}$, and matrices $\YR'$ and $\Ur$ are given, respectively, in eqs.~\eqref{eq:yukawastructures} and~\eqref{eq:mixMR}, the couplings $ \mathbf{\Delta}^k $ in eq.~\eqref{eq:Deltadef} are
\begin{align}
   \; \mathbf{\Delta}^1=\dfrac{M_2}{2 u }\dfrac{\sqrt[4]{r_{12}}}{1+\sqrt{r_{12}}}\begin{pmatrix}
    -2\sqrt[4]{r_{12}}&1-\sqrt{r_{12}}\\
    \cdot&2\sqrt[4]{r_{12}}
    \end{pmatrix}e^{i(\theta_S+\theta)},\quad \mathbf{\Delta}^2=-i\mathbf{\Delta}^1.
    \label{eq:Deltasimp}
\end{align}
Having defined all interactions relevant for the CP asymmetries in $N_{1,2}$ decays, it is worth commenting on some of their properties before proceeding to a detailed numerical analysis of the parameter space:
\begin{itemize}

    \item Plugging $\Y^1$ and $\Y^2$ in the expression for $\varepsilon_{i \alpha}^a(\text{type-I})$ of eq.~\eqref{eq:CPasymtypeI}, one sees that  $\varepsilon_{i \alpha}^a(\text{type-I}) = 0$. In fact, the products $\mathbf{Y}^{a \ast}_{\alpha i} \mathbf{Y}^{b \ast}_{\beta i} \mathbf{Y}^{b}_{\alpha j} \mathbf{Y}^{a}_{\beta j}$, $\mathbf{Y}^{a \ast}_{\alpha i} \mathbf{H}^b_{i j} \mathbf{Y}^{a}_{\alpha j}$ and $\mathbf{Y}^{a \ast}_{\alpha i} \mathbf{H}^b_{j i} \mathbf{Y}^{a}_{\alpha j}$ are real for $i=1,2\neq j$ and $a,b=1,2$. This happens because the matrix entries in each row of $\Y^1$ and $\Y^2$ have the same phase. Thus, we conclude that in our $\mathcal{Z}_8$ model the usual type-I seesaw diagrams of figure~\ref{fig:usualcontribution_CPasym} do not contribute to the CP asymmetry. Consequently, the $(B-L)$-asymmetry will be exclusively generated through the new $h_k$-induced interactions, namely the scalar-fermion portal $N N h_k$ and scalar triple coupling $\Phi^\dagger \Phi h_k$.
    
    \item As for $\varepsilon_{i \alpha}^a(\text{wave})$ of eq.~\eqref{eq:CPasymwave}, using the above expressions for $\mathbf{Y}^{1,2}$ and $\mathbf{\Delta}^{1,2}$, we notice that the products $\mathbf{Y}_{\alpha l}^a \mathbf{\Delta}_{l j}^k \mathbf{\Delta}_{j i}^{k \ast} \mathbf{Y}_{\alpha i}^{a \ast}$ and $\mathbf{Y}_{\alpha l}^a \mathbf{\Delta}_{l j}^{k \ast} \mathbf{\Delta}_{j i}^k \mathbf{Y}_{\alpha i}^{a \ast}$ are real, for any combination of indices $i$, $j$ and $l$. Consequently, the $LL$ and $RR$ contributions to~$\varepsilon_{i \alpha}^a(\text{wave})$ vanish.

    \item Trilinear scalar terms $(\Phi_a^\dagger \Phi_b) S$ as those of eq.~\eqref{eq:cubicscalar} are forbidden by the $\mathcal{Z}_8$ symmetry -- see table~\ref{tab:part&sym}. Hence, the effective coupling $\tilde{\mu}_{a b,k}$ defined in the mass-eigenstate basis as shown in eq.~\eqref{eq:muCPdef}, will be generated via the $\lambda_{1 S,2S} (\Phi_{1,2}^\dagger \Phi_{1,2} ) |S|^2$ quartics, once $S$ aquires a VEV at the leptogenesis scale. Consequently, in our model only $\tilde{\mu}_{a a,k}$ for $a,k =1,2$ will be non-zero, being proportional to $\lambda_{1 S}u$ and $\lambda_{2 S}u$. The $\tilde{\mu}_{ijk}$ couplings of eq.~\eqref{eq:muhkCPdef} will originate from $\lambda_{S} |S|^4$ and $\lambda_{S}^\prime (S^4 + S^{\ast 4})$, being proportional to $ \lambda_{S} u $ and $\lambda_{S}^\prime u $. Moreover, the $\varepsilon_{i \alpha}^a(\text{vertex})$ and $\varepsilon_{i \alpha}^a(\text{3- body decay})$ CP-asymmetries in eqs.~\eqref{eq:CPasymvertex} and~\eqref{eq:CPasym3body} are suppressed by $\tilde{\mu}_{a b, k}/M_i \propto u \lambda/M_i$ if $M_i \lsim u$. This has been  checked numerically and holds even for $\lambda \sim \mathcal{O}(1)$. For example, taking $u/M_i \sim 0.1$ (see figure~\ref{fig:etaBwDeltaL=1}) and $\lambda = 0.01$, we have $|\varepsilon_{i \alpha}^a(\text{wave})| \sim 10^3 \times |\varepsilon_{i \alpha}^a(\text{vertex}) + \varepsilon_{i \alpha}^a(\text{3-body decay})|$.
    
\end{itemize}
In conclusion, the only relevant contribution to the CP asymmetry in the $\mathcal{Z}_8$ model comes from the new wave diagrams~(\subref{fig:newwave}) of figure~\ref{fig:newcontribution_CPasym}, which are possible due to the presence of $S$. Thus, in this simple model, $S$ is responsible for SCPV and for the CP asymmetries in the heavy neutrino decays. Hence, in the present framework, leptogenesis is assisted by $S$ with CP violation coming from its VEV. 

By replacing eqs.~\eqref{eq:Y1},~\eqref{eq:Y2} and~\eqref{eq:Deltasimp} in the general expression for $\varepsilon_{i\alpha}^a(\text{wave})$ in eq.~\eqref{eq:CPasymwave} with $n_H=2$ and $n_S=1$ we have
\begin{align}
   \varepsilon_\text{CP}\equiv\varepsilon_2=& - \dfrac{1}{8\pi}\dfrac{M_2^2}{ u^2}\dfrac{r_{12}\left[1-\sqrt{r_{12}}\right]\left[(1-\sqrt{r_{12}})^2y^2-x^2\sqrt{r_{12}}-xzt_\beta^2\sqrt{r_{12}}\right]}{\left[1+\sqrt{r_{12}}\right]^2\left[r_{12}x^2+(\sqrt{r_{12}}-1)^2y^2+xzt_\beta^2r_{12}\right]}\sin[2(\theta_S+\theta)]\nonumber\\[0.2cm]
    &\times\left[\mathcal{F}_{\text{w}, L R}^{2 1 1 1}-\mathcal{F}_{\text{w}, L R}^{2 1 2 1} -\mathcal{F}_{\text{w}, R L}^{2 1 1 1}+\mathcal{F}_{\text{w}, R L}^{2 1 2 1}\right] \; ,
    \label{eq:CPasymmetrymodel}
\end{align}
after summing over flavour $\alpha=e,\mu,\tau$ and number of Higgs doublets~$a=1,2$ [see eq.~\eqref{eq:cptotunflavoured}]. Here, the doublet VEVs are related as follows,
\begin{equation}
t_\beta \equiv \tan\beta = \frac{v_2}{v_1} \; ,\;
v^2 = v_1^2 + v_2^2=246\,{\rm GeV}\,.
\label{eq:tanbeta}
\end{equation}
Let us highlight some features of the obtained CP asymmetry:
\begin{itemize}

    \item The CP asymmetry generated by the decay of $N_1$ is identically zero, i.e. $\varepsilon_1=0$, due to the kinematic constraint $M_2>M_1+m_{h_k}$ imposed by the decay $N_2\to N_1+h_k$.
    
    \item A direct connection between high and low-energy CP violation is established through the explicit dependence of $\varepsilon_\text{CP}$ on the CP violating phase $\theta$ [recall that $\theta_S$ depends as well on $\theta$ as shown in eqs.~\eqref{eq:thetaS1} and~\eqref{eq:thetaS2}]. As long as $r_{12}\neq 0,1$ and $m_{h_1}/m_{h_2}\neq0,1$, a non-trivial low-energy CP violating phase is required to have non-zero CP asymmetry. This is allowed by neutrino oscillation data (see table~\ref{tab:compatibility}).
    
    \item A negative CP asymmetry, required to ensure $\eta_B>0$ [see eqs.~\eqref{eq:etaB} and \eqref{eq:BEsBL}], is obtained for
\begin{align}
\begin{cases}
    0<r_{12}<{r_{12}}_\text{lim}\\
    \sin[2(\theta_S+\theta)]<0
\end{cases}\quad
\text{ or }\quad\quad
\begin{cases}
    {r_{12}}_\text{lim}<r_{12}<1\\
    \sin[2(\theta_S+\theta)]>0
\end{cases},
\label{eq:negcpasym}
\end{align}
where the limiting $\sqrt{r_{12}}$ value reads
\begin{align}
    \sqrt{{r_{12}}_\text{lim}}=\dfrac{x^2+2y^2+t_\beta^2 x z-\sqrt{(x^2+t_\beta^2 xz)(x^2+4y^2+t_\beta^2 x z)}}{2y^2}.
    \label{eq:r12lim}
\end{align}
In the scenario where $r_{12}=0,1$ one gets $\varepsilon_\text{CP}=0$. 

\item The $\varepsilon_\text{CP}$ dependence on the scalar masses $m_{h_1}$ and $m_{h_2}$ can be analysed from the $LR$ and $RL$ wave loop functions in eq.~\eqref{eq:loopwave} and the $\theta_S$ expression in eq.~\eqref{eq:thetaS2}. When $m_{h_1}/m_{h_2}\rightarrow 0$, $\theta_S \rightarrow - \theta$, and $\varepsilon_\text{CP} \rightarrow 0$. Also, if $m_{h_1}/m_{h_2}\rightarrow1$ then $\mathcal{F}_{\text{w}, L R}^{2 1 1 1}=\mathcal{F}_{\text{w}, L R}^{2 1 2 1}$ and  $\mathcal{F}_{\text{w}, R L}^{2 1 1 1}=\mathcal{F}_{\text{w}, R L}^{2 1 2 1}$ and $\varepsilon_\text{CP}$ vanishes as well. For $0<m_{h_1}/m_{h_2}<|\tan\theta|$ [see eq.~\eqref{eq:mh1mh2ratiolimit}], the CP asymmetry grows with $m_{h_1}/m_{h_2}$ and, thus, one should consider $m_{h_1}/m_{h_2}=|\tan\theta|$ in order to maximize $\varepsilon_\text{CP}$. Furthermore, $|\varepsilon_\text{CP}|$ is enhanced for high $\sigma_{k2}$, $k=1,2$. This can be seen by looking at the $\rho_{ijk}$ dependence on $\sigma_{ki}$ -- see eq.~\eqref{eq:loopwave}. Moreover, $\varepsilon_\text{CP}$ scales with $M_2^2/u^2$.

\end{itemize}

To obtain BAU predictions in the present model, we numerically solve the set of BEs given in~\eqref{eq:BEsNi} and \eqref{eq:BEsBL} for $n_R=2$ and $M_2>M_1$. Namely, 
\begin{align}
    \frac{d N_{N_1}}{d z} & = -(D_1 + \frac{N_{N_2}^{\text{eq}}}{N_{N_1}^{\text{eq}}} D_{2 1} + S_1) (N_{N_1} - N_{N_1}^{\text{eq}}) + D_{2 1} (N_{N_2} - N_{N_2}^{\text{eq}}) \nonumber \\
    & - S_{1 1} [(N_{N_1})^2 - (N_{N_1}^{\text{eq}})^2] - S_{1 2} (N_{N_1} N_{N_2} - N_{N_1}^{\text{eq}} N_{N_2}^{\text{eq}}) \; , \label{eq:BEsN1model} \\
    \frac{d N_{N_2}}{d z} & = - (D_2 + D_{2 1} + S_2)(N_{N_2} - N_{N_2}^{\text{eq}}) + \frac{N_{N_2}^{\text{eq}}}{N_{N_1}^{\text{eq}}} D_{2 1} (N_{N_1} - N_{N_1}^{\text{eq}}) \nonumber \\
    & - S_{2 2} [(N_{N_2})^2 - (N_{N_2}^{\text{eq}})^2] - S_{1 2} (N_{N_1} N_{N_2} - N_{N_1}^{\text{eq}} N_{N_2}^{\text{eq}}) \; , \label{eq:BEsN2model} \\
    \frac{d N_{B-L}}{d z} & = - \varepsilon_2 D_2 (N_{N_2} - N_{N_2}^{\text{eq}}) - W N_{B-L} \; ,
    \label{eq:BEsBLmodel}
\end{align}
which agree with those obtained in refs.~\cite{LeDall:2014too,Alanne:2017sip,Alanne:2018brf}. We will take the values of  $x$, $y$, $z$, $\theta_L$ and $\theta$ fixed to their best-fit values (see table~\ref{tab:compatibilityoutput}) for case $\mathcal{Z}_8^\mu$. Since in our model $\varepsilon_1=0$, $N_{B-L}$ will be exclusively generated by the decay of~$N_2$ into leptons and scalar doublets, when interactions involving $N_2$ go out of equilibrium. The lepton asymmetry is then washed out by the $L$-violating inverse decays and scatterings involving $N_1$ and $N_2$. The resulting baryon-to-photon ratio $\eta_B$ depends on the strength of these washout processes controlled by the interaction couplings and mass ratios of the participating states. After solving the BEs, we take $N_{B-L}(z\rightarrow\infty)$ and use eq.~\eqref{eq:etaB} to compute~$\eta_B$.

In figure~\ref{fig:etaBwDeltaL=1}, we show our baryon-to-photon ratio $\eta_B$ results in the planes $(M_2/u,M_1/M_2\equiv\sqrt{r_{12}})$, $(m_{h_2}/M_2\equiv\sqrt{\sigma_{22}}$, $M_1/M_2=\sqrt{r_{12}})$ and $(t_\beta,M_1/M_2=\sqrt{r_{12}})$ on the upper-left, upper-right and bottom figures, respectively. In the upper-left (upper-right) [bottom] plot we considered $m_{h_2}/M_2=0.1$ and $t_\beta=3$ ($y_{R_S}=4\pi$ and $t_\beta=3$) [$m_{h_2}/M_2=0.1$ and $y_{R_S}=4\pi$]. In all cases we set $v=246$~GeV [see eq.~\eqref{eq:tanbeta} for relation with $t_\beta$], $m_{h_1}/m_{h_2}=0.37$ and $u=10^{12}$~GeV (unflavoured scenario). The parameter space regions where the kinematic constraint $M_2>M_1+m_{h_2}$ is not met are depicted in gray. Also, the yellow regions in the upper-right and bottom plots indicate parameter space where $\eta_B<0$ (corresponding to $\sqrt{r_{12}}\gsim 0.94$ for $t_\beta=3$ in this benchmark). Additionally, we imposed the perturbativity constraints $|\lambda_i| \leq 4 \pi$ and $|y_{D_i}|,|y_{R_S}| \leq 4 \pi$, and required the scalar potential to be bounded from below [see eq.~\eqref{eq:bfb}]. For the chosen benchmark, among these conditions only $y_{R_S}$ perturbativity fails, which is indicated by the black hatched area in the upper-left plot. For the remaining figures $y_{R_S}=4\pi$ is imposed to maximise the ratio $M_2/u$, which in turn maximises $\eta_\text{B}$. The blue contour regions show the $\eta_B$ results obtained when considering only the contribution of decays and inverse decays to the final asymmetry. The black dashed contours correspond to $\eta_B$ values computed considering $\Delta L=1$ and $\Delta N=2$ scattering terms in the BEs. The red dashed (solid) line indicates the contour for $\eta_B=\eta_B^0$ -- see eq.~\eqref{eq:etab0} -- when scattering effects are (not) included in the BEs.

From the results shown in that figure we conclude the following:
\begin{itemize}
\item From the upper-left plot one can see that, for the considered benchmark with $t_\beta=3$ and $m_{h_2}/M_2=0.1$, the experimental value of the BAU $\eta_B^0$ is obtained for $M_1\gtrsim 0.81\, M_2$ ($M_1\gtrsim 0.86\, M_2$) and $M_2\gtrsim 3.5\, u$ ($M_2\gtrsim 5.2\, u$), when decays and inverse decays (decays, inverse decays and scatterings) are included in the BEs. In the upper-right plot, for which $y_{R_S}=4\pi$, $\eta_B=\eta_B^0$ for $M_1\gtrsim 0.69\, M_2$ ($M_1\gtrsim 0.71 \,M_2$) and $m_{h_2}\gtrsim 4\times 10^{-2} M_2$ ($m_{h_2}\gtrsim 7\times 10^{-2} M_2$). Choosing $m_{h_2}/M_2=0.1$ and keeping $y_{R_S}=4\pi$ (bottom plot), the observed value for the baryon-to-photon ratio lies in the region $M_1\gtrsim 0.81\, M_2$ ($M_1\gtrsim 0.86\, M_2$) and $t_\beta\lesssim 8.4$ ($t_\beta\lesssim 5.6$). The fact that $\eta_\text{B}^0$ can only be recovered in a small portion of the parameter space, even though $\varepsilon_\text{CP} \sim [10^{-5},10^{-2}]$, is due to the very strong washout regime (with $K_i\gsim 2\times 10^2$). Thus, only for high $M_1/M_2$, $M_2/u$, and $m_{h_2}/M_2$, where $\varepsilon_\text{CP}$ is maximised, ($B-L$)-asymmetry is sufficiently high. As seen in the upper-right plot of figure~\ref{fig:etaBwDeltaL=1}, the lower limit on~$M_1/M_2$ decreases when the ratio $m_{h_2}/M_2$ increases. Moreover, by looking at the yellow region in the bottom plot we notice that as $t_\beta$ increases the available parameter space where the baryon-to-photon ratio is positive shrinks. In fact, since the SCPV phase is predicted to be $\theta\simeq 1.89\pi$ we have $\sin[2(\theta_S+\theta)]<0$ ($\sim-0.54$ for $m_{h_1}/m_{h_2}=|\tan\theta|\simeq0.37$), which allows for $\eta_B>0$ when $0<r_{12}<{r_{12}}_\text{lim}$ [see eq.~\eqref{eq:negcpasym}]. For low values of $t_\beta$, ${r_{12}}_\text{lim}\simeq 1$ and, consequently, the available parameter space for successful leptogenesis is larger.

\item Some of the scattering processes presented in section~\ref{sec:BEs} (for which the reduced cross section are given in appendix~\ref{sec:scatterings}) have negligible effect on the evolution of $N_i$ number density and ($B-L$)-asymmetry. In fact, we numerically checked that in our model the $\Delta N=2$ neutrino annihilation scatterings (see figure~\ref{fig:DeltaNeq2_scattering}) are subdominant compared to the $\Delta L=1$ scatterings (see figure~\ref{fig:DeltaLeq1_scattering}). As mentioned before, in our scenario the effective triple scalar coupling $\tilde{\mu}_{a b, k} \propto u \lambda$ is naturally suppressed compared to the heavy neutrino masses~$M_{1,2}$. Hence, among the different contributions to $\Delta L=1$ scatterings, the ones involving this triple scalar interaction mediated by $\Phi_b$ will be subdominant compared to the corresponding $N$-mediated one (this has been verified numerically). Thus, among $\Delta L=1$ scatterings, the dominant ones are: the usual type-I seesaw diagrams of figure~\ref{fig:DeltaLeq1_usual}, the heavy neutrino mediated $t$-channel $N_i h_k \leftrightarrow \Phi_a \ell_\alpha$ and $N_i \Phi_a \leftrightarrow \ell_\alpha  h_k$ processes shown in figures~\ref{fig:DeltaLeq1_Nltophih} and \ref{fig:DeltaLeq1_Nphitohl}, respectively, and the $s$-channel contribution to $N_i  \ell_\alpha \leftrightarrow \Phi_a h_k$ in figure~\ref{fig:DeltaLeq1_Nhtophil}.

\item From the upper-left plot in figure~\ref{fig:etaBwDeltaL=1}, we distinguish two different scenarios: for $M_2/u\lsim 3$ scatterings are negligible while for $M_2/u\gsim3$ they significantly lower the value of $\eta_B$. In the former case, the dominant scattering effects stem from the regular type-I seesaw contributions of figure~\ref{fig:DeltaLeq1_usual}, which scale as $\mathcal{O}(Y^2Y_t^2)$. For low values of $M_2/u$, we verify that $\mathcal{O}(\Delta)\ll\mathcal{O}(Y_t)$ and, thus, the new scattering contributions of order $\mathcal{O}(Y^2\Delta^2)$ are subdominant. However, due to compatibility with neutrino data, we are in a strong washout regime with $K_i\gsim 2\times 10^2$, and the impact of the usual top-quark scatterings is negligible (as already remarked in refs.~\cite{Buchmuller:2004nz,Hahn-Woernle:2009jyb}). Instead, for $M_2/u\gsim3$, the scatterings with the new scalars $h_k$ are significantly enhanced [note the $\mathbf{\Delta}_{1,2}$ dependence on this ratio in eq.~\eqref{eq:Deltasimp}], becoming out of equilibrium much later in the early Universe. This leads to a stronger washout of the $(B-L)$-asymmetry and, consequently, to a lower $\eta_\text{B}$. In this region of the parameter space, we checked that the dominant scattering contribution corresponds to the $N$-mediated $s$-channel process $N h \leftrightarrow  \ell \Phi$. The dominance of this process over the usual $\Delta L=1$ type-I seesaw contributions is explained by the fact that $\mathcal{O}(Y_t)\ll \mathcal{O}(\Delta)$. Moreover, the new $t$-channel contributions $N_i \ell_\alpha \leftrightarrow \Phi_a h_k$ and $N_i \Phi_a \leftrightarrow \ell_\alpha h_k$ mediated by $N$ are naturally subdominant when compared to the (also $N$-mediated) $s$-channel process $N_i h_k \leftrightarrow \ell_\alpha \Phi_a$, due to the logarithmic dependence on the mediator mass.

\item From eq.~\eqref{eq:CPasymmetrymodel}, it is clear that $\varepsilon_\text{CP}$ strongly depends on the SCPV phase $\theta$ through the factor $\sin[2(\theta+\theta_S)]$. By varying $\theta$ in the allowed 3$\sigma$ range for neutrino oscillation data we get $\theta\sim [1.89,1.96]\,\pi$, being the best-fit point $\theta=1.89 \pi$ (see table~\ref{tab:compatibilityoutput}). Increasing $\theta$ up to its largest allowed value, leads to a lower $|\varepsilon_\text{CP}|$, and consequently, to a lower BAU. Hence, using the best-fit value for the vacuum CP phase in our analysis maximises $\eta_\text{B}$.

\end{itemize}

To finalise our discussion, a few comments are in order regarding charged-lepton flavour violation~(cLFV). At low-energies, i.e. at the EW scale, the two Higgs doublets $\Phi_{1,2}$ will develop non-zero VEVs $v_{1,2}$, as indicated in eq.~\eqref{eq:vevs}. The angle $\beta$ [see eq.~\eqref{eq:tanbeta}] diagonalises the charged scalar mass matrix leading to the $W$-type Goldstone boson and a charged Higgs $H^{\pm}$. Furthermore, $\beta$ defines the Higgs basis where $\Phi_{1}$ matches the SM Higgs doublet~\cite{Branco:2011iw}. Note that, since $u \gg v_{1,2}$, the singlet~$S$ is decoupled from the doublets $\Phi_{1,2}$. So, we work in the alignment limit where we set $\beta-\alpha = \pi/2$, being $\alpha$ is the angle which which rotates the neutral doublet degrees of freedom to their mass basis [see eq.~\eqref{eq:scalardef}]. The new scalars will mediate the cLFV decays $\ell_{\alpha}^{-} \rightarrow \ell_{\beta}^{-} \ell_{\gamma}^{+} \ell_{\delta}^{-}$ and $\ell_{\alpha} \rightarrow \ell_{\beta} \gamma$ at tree and one-loop level, respectively. For the $\mathcal{Z}_8^\mu$ case discussed above, all contributions to the $\mu \rightarrow 3 e$, $\mu \rightarrow e \gamma$ and $\mu-e$ conversion in nuclei, vanish. This is due to the presence of zeros imposed by the flavour symmetry on the charged lepton Yukawa matrices~$\Yl^{1,2}$ of eq.~\eqref{eq:yukawastructures} and, consequently, on the mass matrix $\Ml$ shown in eq.~\eqref{eq:masstructures} -- see refs.~\cite{Correia:2019vbn,Camara:2020efq}. For this $\mu$-decoupled case, the neutral scalars only contribute to~$\tau \rightarrow 3 e$ and $\tau \rightarrow e \gamma$, whose current bounds are orders of magnitude above the stringent muon cLFV ones. Naturally, these contributions are suppressed for large scalar masses or if they are quasi-degenerate. Lastly, the $W^{\pm}$ and  $H^{\pm}$ will also contribute radiatively to the aforementioned cLFV processes, due to their interactions with Majorana neutrinos. For low-scale seesaw scenarios where heavy neutrinos can have masses of the order~$\mathcal{O}(1\,{\rm TeV})$, these contributions are testable at current and future indirect cLFV experiments, as studied e.g. in ref.~\cite{Camara:2020efq} for the minimal inverse-seesaw. However, for the canonical type-I scenario analysed in this paper, the heavy neutrino masses are around $ 10^{12}$ GeV and, consequently, the contributions to cLFV from charged bosons are naturally suppressed.

\section{Concluding remarks}
\label{sec:concl}

In this paper we explored thermal leptogenesis within the canonical type-I seesaw model extended with complex scalar singlets. Provided that CP invariance is imposed at the Lagrangian level, the complex VEVs of scalar singlets will be the unique source of both Dirac and Majorana CP violation, at the EW scale, and high-energy CP violation at the leptogenesis scale. These scalars unlock novel radiative corrections to the CP asymmetry generated when the heavy neutrinos decay into leptons, and provide new tree-level CP-violating three-body decay processes. In this work, we generalised the CP asymmetry calculation for an arbitrary number of RH neutrinos, complex scalar singlets and Higgs doublets, as shown in eq.~\eqref{eq:fullCP}. Furthermore, we studied the unflavoured BEs taking into account decays, $\Delta L=1$ and $\Delta N =2$ scatterings for an arbitrary number of RH neutrinos. The new complex scalar singlets participate in additional tree-level decays and scattering processes besides the ones considered in vanilla type-I seesaw leptogenesis. In order to compute the final values of the baryon-to-photon ratio, one has to solve the BEs presented in  eqs.~\eqref{eq:BEsNi} and \eqref{eq:BEsBL}. The reduced cross-sections for all included processes in the BEs are collected in appendix~\ref{sec:scatterings}.

To illustrate how SCPV can simultaneously lead to non-trivial low- and high-energy CP-violation, we studied a simple model where the SM is extended with one complex singlet, two RH neutrinos and a new scalar doublet. The parameters in the Lagrangian are further constrained by a $\mathcal{Z}_8$ flavour symmetry. This corresponds to the minimal particle content charged under the simplest discrete symmetry that allows for the possibility of SCPV and compatibility with neutrino oscillation data. The $\mathcal{Z}_8$ leads to constraints in the effective neutrino mass matrix which we tested against data. We concluded that out of the three possible charge assignments, $\mathcal{Z}_8^e$, $\mathcal{Z}_8^\mu$ and $\mathcal{Z}_8^\tau$, and considering all possible decoupled charged-lepton states ($e$, $\mu$ or $\tau$), the best case is the $\mathcal{Z}_8^{\mu}$ with muon decoupled and IO neutrino masses (see tables~\ref{tab:compatibility} and~\ref{tab:compatibilityoutput}), which requires a vacuum singlet phase $\theta\sim 1.89\pi$. 

Due to the constrained structure of the Yukawa couplings $\mathbf{Y}^{1,2}$ and the couplings of heavy neutrinos to the new scalar singlets, $\mathbf{\Delta}^{1,2}$, the CP asymmetries in the $N_2$ decays stem from the interference between the tree-level and the one-loop self-energy diagrams mediated by the $S$ singlet (the usual type-I seesaw diagrams are forbidden and the new 3-body decay and vertex contributions negligible). The expression for the total CP asymmetry in this model is presented in eq.~\eqref{eq:CPasymmetrymodel}, where the link between low and high-energy CP violation is explicit. To compute the value of the BAU, we solved numerically the unflavoured BEs for low-energy parameters that best fit neutrino data, including scattering processes. We concluded that for case $\mathcal{Z}_8^\mu$ successful leptogenesis is achieved when $M_2 \gsim 8\, u$, for $u=10^{12}$ GeV, as shown in the upper-left plot of figure~\ref{fig:etaBwDeltaL=1}. The exact $M_2/u$ lower bound increases slightly when scatterings are included, thanks to the new washout sources. Furthermore, we showed that scatterings are only relevant for $M_2\gsim 3\,u$, lowering significantly $\eta_\text{B}$. The values of $M_1/M_2$ above which the observed BAU is recovered strongly depend on $m_{h_2}/M_2$ and $t_\beta=v_2/v_1$, as shown in the upper-right and bottom plots of figure~\ref{fig:etaBwDeltaL=1}. Thus, we conclude that case $\mathcal{Z}_8^\mu$ recovers the observed BAU for part of the parameter space mainly depending on the ratios $M_1/M_2$, $m_{h_2}/M_2$ and $M_2/u$.

The general setup of scalar-singlet assisted leptogenesis with SCPV discussed here can be applied in a straightforward way to the type-III seesaw framework, as well as to the canonical scotogenic model. The Majorana mass term provides the link between SCPV induced by the scalar singlet VEVs and LCPV. Regarding leptogenesis, the expressions for the CP-asymmetry and BEs are essentially the same as the ones obtained here. The generalisation for the type-II seesaw case can also be done, but the CP-asymmetries, BEs and portal linking SCPV and the neutrino sector are distinct. This interesting possiblity will be explored in a future work. 

\acknowledgments
We are grateful to T. Alanne and A. Ritz for private communications regarding details on some cross-section results. This work is supported by Fundação para a Ciência e a Tecnologia (FCT, Portugal) through the projects CFTP-FCT Unit UIDB/00777/2020 and UIDP/00777/2020, CERN/FIS-PAR/0004/2019, which are partially funded through POCTI (FEDER), COMPETE, QREN and EU. The work of D.B. and H.C. is supported by the PhD FCT grants SFRH/BD/137127/2018 and 2021.06340.BD, respectively. FRJ thanks the CERN Department for Theoretical Physics for hospitality and financial support during the final stage of this work.

\appendix

\section{General aspects of Boltzmann equations}
\label{sec:genBEs}

In this section we collect the expressions, formulas and notation we use in this work to write down the BEs in section~\ref{sec:BEs}, following refs.~\cite{Buchmuller:2000as,Buchmuller:2002rq,Buchmuller:2004tu,LeDall:2014too}. We work with the relativistic formulation of classical BEs in the Friedman-Robertson-Walker metric, where we assume the quantum coherence effects to be negligible. For recent examples where a quantum treatment of BEs is reviewed see refs.~\cite{Biondini:2017rpb,Garbrecht:2018mrp,Klaric:2021cpi}.

In the early Universe interactions among particles in the thermal bath and the expansion of the Universe influence the microscopic time evolution of particle number densities and asymmetries which is described by a coupled system of BEs. Considering the number density~$n_\psi$ of a particle species $\psi$, the BEs take the following form~\cite{Kolb:1990vq}
    \begin{align}
    \frac{d n_\psi}{d t} + 3 H n_\psi & = - \sum_{i, j, \cdots} \left[ \gamma\left( \psi \rightarrow i + j + \cdots \right) - \gamma\left( i + j + \cdots \rightarrow \psi \right) \right] \nonumber \\
    & - \sum_{a, i, j, \cdots} \left[\gamma\left(\psi + a \rightarrow i + j + \cdots\right) - \gamma\left(i + j + \cdots \rightarrow \psi + a\right)\right] \; .
    \label{eq:BEsgeneric1}
    \end{align}
    In the above equation, the left-hand side takes into account effects of the expansion of the Universe, while the right-hand side is the collision term involving interactions. The number density and Hubble parameter are given by,
    \begin{align}
    n_\psi = \frac{g_\psi}{(2 \pi)^3} \int d^3 p_\psi f_\psi\; ,\quad \; H(T) = \sqrt{\frac{4 \pi^3 g_{\ast}}{45}} \frac{T^2}{M_{\text{Pl}}} \; ,
    \label{eq:Hubble}
    \end{align}
    where $g_\psi$ and $f_\psi$ are the number of internal degrees of freedom and phase space distribution of the particle species $\psi$. For example, $g_{N} = 2$ for Majorana neutrinos, $g_\ell = 2$ for lepton doublet components, $g_{e_R}=1$ for charged-lepton singlet fields and $g_\Phi=2$ for Higgs doublet components. Furthermore, $g_{\ast}$ is the effective number of relativistic degrees of freedom in the thermal bath at temperature $T$. In the limit of very high temperatures, we have $g_{\ast} = 106.75$ in the SM case. Finally, $M_{\text{Pl}}\simeq 1.22 \times 10^{19}$~GeV is the Planck mass. For a general process $\gamma_{\psi + a + b + \cdots  \rightarrow i + j + \cdots} \equiv \gamma\left(\psi + a + b + \cdots  \rightarrow i + j + \cdots\right)$, involving $\psi$, the collision term is
     \begin{align}
    \gamma_{\psi + a + b + \cdots  \rightarrow i + j + \cdots} & = \int \frac{d^3 p_\psi}{(2 \pi)^3 2 E_\psi} \frac{d^3 p_a}{(2 \pi)^3 2 E_a} \frac{d^3 p_b}{(2 \pi)^3 2 E_b} \cdots \frac{d^3 p_i}{(2 \pi)^3 2 E_i} \frac{d^3 p_j}{(2 \pi)^3 2 E_j} \cdots  \label{eq:collision} \\ 
    & \times (2 \pi)^4 \delta^4\left(p_\psi + p_a + p_b + \cdots  \rightarrow p_i + p_j + \cdots\right)  \nonumber \\ 
    & \times \left|\mathcal{M}\left(\psi + a + b + \cdots  \rightarrow i + j + \cdots\right) \right|^2 f_\psi f_a f_b \cdots (1 \pm f_i) (1 \pm f_j) \; , \nonumber
    \end{align}
    with the phase space integrals containing $p_\psi$ and $E_\psi$, being the momentum and energy of a given particle $\psi$ with mass $m_\psi$. In the above, the Dirac $\delta$-function accounts for momentum conservation and the squared S-matrix element $\left|\mathcal{M}\left(\psi + a + b + \cdots  \rightarrow i + j + \cdots\right) \right|^2$ is summed over the internal degrees of freedom of incoming and outgoing particles taking into consideration appropriate symmetry factors. Furthermore, the upper (lower) sign in $(1 \pm f_i)$ refers to bosons (fermions).
    
    Working in the dilute gas approximation we may consider $(1 \pm f_i) \simeq 1$. Furthermore, elastic scatterings will only affect the phase space densities of particles while inelastic scatterings change their number densities. Assuming that the elastic scatterings are fast enough to maintain kinetic equilibrium, in comparison to the inelastic ones, the phase space and number densities are related through $f_\psi(E_\psi, T) = n_\psi e^{- E_\psi / T}/n_\psi^{\text{eq}}$, where we use the Maxwell-Boltzmann equilibrium distribution. We have,
    \begin{align}
    n_{\psi}^{\text{eq}} = g_\psi \frac{m_\psi^2}{2 \pi^2} T \; \mathcal{K}_2\left(\frac{m_\psi}{T}\right) \; , \; n_{\gamma}^{\text{eq}} = g_\gamma \frac{T^3}{\pi^2} \; ,
    \label{eq:equi}
    \end{align}
    with $\mathcal{K}_n(x)$ being the modified Bessel function of order $n$. To automatically take into account effects of the expansion of the Universe, it is convenient to work with the particle number $N_\psi$ in the comoving volume $R_\ast(t)^3 = n_\gamma^{\text{eq}}(t)^{-1}$, which contains one photon at time $t_\ast$ before leptogenesis takes place~\cite{Buchmuller:2002rq}. Namely, we use the following variables,\footnote{In the literature is often used $Y_\psi = n_\psi/s$ which normalizes the particle number density to the entropy~$s$. In an isentropically expanding Universe (entropy is conserved), $N_\psi$ and $Y_\psi$ are related by a constant.}
    \begin{align}
    z_\psi=\frac{m_\psi}{T} \; , \; N_\psi = \frac{n_\psi}{n_\gamma^{\text{eq}}} \; , \; N_{\psi}^{\text{eq}} = \frac{n_{\psi}^{\text{eq}}}{n_\gamma^{\text{eq}}} \; , \; N_{N}^{\text{eq}} = \frac{3}{8} z_{N}^2 \mathcal{K}_2(z_{N}) \; , \; N_{\ell}^{\text{eq}} = \frac{3}{4}\; ,
    \label{eq:ournotation}
    \end{align}
    where we explicitly write the equilibrium particle number for Majorana neutrinos and SU(2) lepton-doublet components which are used in this work (see section~\ref{sec:BEs}). Under the stated assumptions and performing the above change of variables the BEs in eq.~\eqref{eq:BEsgeneric1} become,
    \begin{align}
    n_\gamma^{\text{eq}} z_\psi H(z_\psi) \frac{d N_\psi}{d z_\psi} & = - \sum_{i, j, \cdots} \left[\frac{N_\psi}{N_\psi^{\text{eq}}} \gamma^{\text{eq}}_{\psi \rightarrow i + j + \cdots } - \frac{N_i N_j \cdots}{N_i^{\text{eq}} N_j^{\text{eq}} \cdots} \gamma^{\text{eq}}_{i + j + \cdots \rightarrow \psi} \right] \nonumber \\
    & - \sum_{a, i, j, \cdots} \left[\frac{N_\psi N_a}{N_\psi^{\text{eq}} N_a^{\text{eq}}} \gamma^{\text{eq}}_{\psi + a \rightarrow i + j + \cdots} - \frac{N_i N_j \cdots}{N_i^{\text{eq}} N_j^{\text{eq}} \cdots} \gamma^{\text{eq}}_{i + j + \cdots \rightarrow \psi + a}\right] \; ,
    \label{eq:BEsgeneric2}
    \end{align}
    with $H(z_\psi) \equiv H(T= m_\psi/z)$ [see eq.~\eqref{eq:Hubble}]. In a dilute gas one only considers decays and two particle scatterings, as well as their back reactions. For the decay we have,
    \begin{align}
         \gamma^{\text{eq}}_{\psi \rightarrow i + j + \cdots} = n_\psi^{\text{eq}} \Gamma_{\psi \rightarrow i + j + \cdots} \frac{\mathcal{K}_1(z_\psi)}{\mathcal{K}_2(z_\psi)} \; ,
         \label{eq:reacdecay}
     \end{align}
     where $\Gamma_{\psi \rightarrow i + j + \cdots}$ is the decay rate of the process $\psi \rightarrow i + j + \cdots$ calculated in the center of mass frame of particle $\psi$. The reaction density for a two-body scattering is given by,
     \begin{align}
         \gamma^{\text{eq}}_{\psi + a \rightarrow i + j + \cdots} = \frac{T}{64 \pi^4} \int_{(m_\psi + m_a)^2}^{\infty} ds \; \hat{\sigma}(s)  \sqrt{s} \mathcal{K}_1\left(\frac{\sqrt{s}}{T}\right) \; ,
         \label{eq:reacscattering}
     \end{align}
    where $s$ is the squared centre-of mass energy and $\hat{\sigma}(s)$ is the reduced cross section for the process $\psi + a \rightarrow i + j + \cdots$. The latter is related to the usual total cross section $\sigma(s)$ through,
    \begin{align}
    \hat{\sigma}(s) = \frac{8}{s} \left[ (p_\psi\cdot p_a)^2 - m_\psi^2 m_a^2 \right] \sigma(s) \; .
    \label{eq:reducedsigma}
    \end{align}

    It is useful to define the following decay and scattering variables, which are just a rescaled version of the quantities in eqs.~\eqref{eq:reacdecay} and~\eqref{eq:reacscattering}, respectively given by
    \begin{align}
         D_{\psi \rightarrow i + j + \cdots } &= \frac{\gamma^{\text{eq}}_{\psi \rightarrow i + j + \cdots}}{n_\gamma^{\text{eq}} N_\psi^\text{eq} z_\psi H(z_\psi)} = K_{\psi \rightarrow i + j + \cdots} z_\psi \frac{\mathcal{K}_1(z_\psi)}{\mathcal{K}_2(z_\psi)}\; , \nonumber \\ 
          K_{\psi \rightarrow i + j + \cdots} &= \frac{\Gamma_{\psi \rightarrow i + j + \cdots}}{H(T=m_\psi)} \; ,
         \label{eq:Decaygen}
    \end{align}
   and,
    \begin{align}
        S_{\psi + a \rightarrow i + j + \cdots} = \frac{\gamma^{\text{eq}}_{\psi + a \rightarrow i + j + \cdots}}{\left(n_\gamma^{\text{eq}}\right)^2 N_\psi^\text{eq} N_a^\text{eq} z_\psi H(z_\psi)} \; .
        \label{eq:Scatteringgen}
    \end{align}
    Neglecting the CP-violating effects and under the assumption of CPT symmetry, energy conservation implies that $\gamma^{\text{eq}}_{\psi + a \rightarrow i + j + \cdots} = \gamma^{\text{eq}}_{i + j + \cdots\rightarrow \psi + a }$. Hence, the inverse-decay parameter is related to the decay variable defined above, as follows,
    \begin{align}
         {ID}_{\psi \rightarrow i + j + \cdots } =  \frac{N_\psi^\text{eq}}{N_i^\text{eq} N_j^\text{eq}} D_{\psi \rightarrow i + j + \cdots } \; .
         \label{eq:InvDecaygen}
    \end{align}

\section{Scattering cross sections}
\label{sec:scatterings}

Here we collect the expressions of the reduced cross sections for the $\Delta L = 1$ (figure~\ref{fig:DeltaLeq1_scattering}) and~$\Delta N = 2$ (figure~\ref{fig:DeltaNeq2_scattering}) two-body scattering processes included in our analysis of the BEs in section~\ref{sec:BEs} and in our numerical computations of section~\ref{sec:BAUmodel}.

  In figure~\ref{fig:DeltaLeq1_scattering}, diagrams (\subref{fig:DeltaLeq1_usual}) are the usual ones occurring in type-I seesaw leptogenesis. The reduced cross-section for the $s$-channel mediated $N_i \ell_{\alpha} \to q_L u_R$ process is given by,
    \begin{align}
    \hat{\sigma}_s(N_i \ell_{\alpha} \rightarrow {q_L}_\beta {u_R}_\gamma)&=  \dfrac{1}{16\pi}\dfrac{\left(s-M_i^2\right)^2}{s^2}\sum_{a,b=1}^{n_H}\Y^{a\ast }_{\alpha i}\Y^b_{\alpha i} (\Y^a_{u})_{\beta \gamma}(\Y^{b\ast}_{u})_{\beta \gamma}\; ,
    \end{align}
    and for the $t$-channel mediated $N_i u_R  \to \ell_{\alpha} q_L$ and $N_i q_L \to \ell_{\alpha} u_R$ processes we have,
    \begin{align}
    \hat{\sigma}_t(N_i {u_R}_\gamma  \rightarrow \ell_{\alpha} q_L) &= \dfrac{1}{16\pi}\dfrac{s-M_i^2}{s}\sum_{a,b=1}^{n_H}\Y^{a }_{\alpha i}\Y^{b\ast}_{\alpha i} (\Y^a_{u})_{\beta \gamma}(\Y^{b\ast}_{u})_{\beta \gamma} \nonumber\\
    & \times \left[1-\dfrac{m_{\Phi_a}^2(M_i^2-m_{\Phi_a}^2)}{(m_{\Phi_b}^2-m_{\Phi_a}^2)(s-M_i^2)} \log\left(\dfrac{s-M_i^2+m_{\Phi_a}^2}{m_{\Phi_a}^2}\right)\right. \nonumber\\
    &\left.+\dfrac{m_{\Phi_b}^2(M_i^2-m_{\Phi_b}^2)}{(m_{\Phi_b}^2-m_{\Phi_a}^2)(s-M_i^2)}\log\left(\dfrac{s-M_i^2+m_{\Phi_b}^2}{m_{\Phi_b}^2}\right)\right]
    \; , \label{eq:Higgsthermal} \\
    \hat{\sigma}_t(N_i q_L \to \ell_{\alpha} {u_R}_\gamma)&=\hat{\sigma}_t(N_i {u_R}_\gamma  \rightarrow \ell_{\alpha} q_L) \, .
    \end{align}
    The expressions above are written in terms of an arbitrary number of Higgs doublets and generic quark Yukawa matrices $\mathbf{Y}_u$. In this work, we assume that all quarks couple diagonally to the first Higgs doublet, which in the alignment limit corresponds to the SM Higgs doublet with VEV equal to $v \simeq 246$ GeV. Furthermore, we only included the dominant top-quark contribution to the above scattering processes in our numerical analysis. The $t$-channel diagrams are mediated by the Higgs and diverge in the limit where we neglect the Higgs masses $m_{\Phi_{a,b}} = 0$. Hence, as commonly done in the literature~\cite{Plumacher:1996kc,Plumacher:1997ru,Plumacher:1998ex,Luty:1992un}, we introduce a Higgs mass $m_{\Phi_{a,b}}/M_i = 10^{-5}$~\cite{Buchmuller:2004nz,LeDall:2014too,Hahn-Woernle:2009jyb}. As previously noted the numerical results are not affected in a substantial way by the chosen prescription.
    
    In figure~\ref{fig:DeltaLeq1_scattering}, the new scattering labelled (\subref{fig:DeltaLeq1_Nltophih}) corresponds to the $N_i \ell_{\alpha} \to \Phi_a h_k$ process, which has the following reduced cross sections,
    \begin{align}
    &\hat{\sigma}_s(N_i \ell_{\alpha} \rightarrow \Phi_a h_k) = \sum_{b,c=1}^{n_H} \frac{\mathbf{Y}_{\alpha i}^{b}\mathbf{Y}_{\alpha i}^{c*} \tilde{\mu}_{a b, k} \tilde{\mu}_{a c, k}^\ast  }{32 \pi} \frac{(s - m_{h_k}^2) (s-M_i^2)^2}{s^4} \; ,
    \end{align}
    \begin{align}
    \hat{\sigma}_t(N_i \ell_{\alpha} \rightarrow \Phi_a h_k) &=\sum_{j,l=1}^{n_R} \frac{(1+\delta_{ij})(1+\delta_{il})|\mathbf{Y}_{\alpha i}^a|^2}{32\pi s^2(M_j^2-M_l^2)}\left\{sM_i(M_j\mathbf{\Delta}_{ij}^k\mathbf{\Delta}_{il}^k+M_l\mathbf{\Delta}_{ij}^{k*}\mathbf{\Delta}_{il}^{k*})\right. \nonumber \\
    &\left.\times\left(M_j^2\log\left[\dfrac{(s-M_i^2)(s-m_{h_k}^2)+sM_j^2}{sM_j^2}\right]-(j\leftrightarrow l)\right)\right.\nonumber\\
    &\left.+sM_jM_l(s-M_i^2)\mathbf{\Delta}_{ij}^k\mathbf{\Delta}_{il}^{k*}\log\left[\dfrac{(s-M_i^2)(s-m_{h_k}^2)M_j^2+sM_j^2M_l^2}{(s-M_i^2)(s-m_{h_k}^2)M_l^2+sM_j^2M_l^2}\right] \right.\nonumber\\ 
    &\left.  -\mathbf{\Delta}_{ij}^{k*}\mathbf{\Delta}_{il}^{k}(M_j^2-M_l^2)(s-M_i^2)(s-m_{h_k}^2) \right.\nonumber\\ 
    &\left. +\mathbf{\Delta}_{ij}^{k*}\mathbf{\Delta}_{il}^{k}\Big( sM_j^2(s+M_j^2-m_{h_k}^2)\right.\nonumber\\ 
    &\left.\times\log\left[\dfrac{(s-M_i^2)(s-m_{h_k}^2)+sM_j^2}{sM_j^2}\right]-(j\leftrightarrow l)\Big)\right\} ,
    \end{align}
    \begin{align}
    \hat{\sigma}_{s-t}(N_i \ell_{\alpha} \rightarrow \Phi_a h_k) &= \sum_{b=1}^{n_H}\sum_{j=1}^{n_R}\dfrac{(1+\delta_{ij})}{16\pi s^3} \Big\{sM_j  \text{Re}\left[\mathbf{Y}_{\alpha i}^b\mathbf{Y}_{\alpha i}^{a*}\tilde{\mu}_{ab,k}(M_iM_j\mathbf{\Delta}_{ij}^{k}-(s-M_i^2)\mathbf{\Delta}_{ij}^{k*})\right]\nonumber\\
    &\times \log\left[\dfrac{(s-M_i^2)(s-m_{h_k}^2)+sM_j^2}{sM_j^2}\right] \nonumber\\
    &-M_i(s-M_i^2)(s-m_{h_k}^2)\text{Re}\left[\mathbf{Y}_{\alpha i}^b\mathbf{Y}_{\alpha i}^{a*}\tilde{\mu}_{ab,k}\mathbf{\Delta}_{ij}^{k}\right]\Big\},
    \end{align}
    which refer to the $s$ and $t$-channel contributions, as well as their interference. The above expression for the $s$-channel contribution matches the result presented in ref.~\cite{LeDall:2014too}. In case the scalar potential parameter $\mu$ is present in the Lagrangian, the $s$-channel diagram is dominant when compared to the $t$-channel one, due to the usual logarithmic dependence with the mediator mass obtained in $t$-channel cross sections~\cite{LeDall:2014too,Alanne:2017sip,Alanne:2018brf}.
    
    Furthermore, diagrams (\subref{fig:DeltaLeq1_Nhtophil}) for $N_i h_k \to \ell_{\alpha} \Phi_a$ lead to,
    \begin{align}
     \hat{\sigma}_s(N_i h_k \rightarrow \ell_{\alpha} \Phi_a) &= \sum_{j,l=1}^{n_R} \frac{(1+\delta_{i j})(1+\delta_{i l})\mathbf{Y}_{\alpha l}^{a}\mathbf{Y}_{\alpha j}^{a*}}{32 \pi} \frac{\sqrt{\rho(s,M_i^2,m_{h_k}^2)}}{s (s-M_j^2)(s-M_l^2)} \nonumber \\
    &  \times \Big[(s+M_i^2-m_{h_k}^2)   \left(s \mathbf{\Delta}^k_{i j} \mathbf{\Delta}^{k \ast}_{i l}  + M_j M_l \mathbf{\Delta}^{k \ast}_{i j} \mathbf{\Delta}^k_{i l}\right) \nonumber \\
    & + 2 s M_i  \left(M_l \mathbf{\Delta}^k_{i j} \mathbf{\Delta}^k_{i l}+M_j \mathbf{\Delta}^{k*}_{i j} \mathbf{\Delta}^{k*}_{i l}\right) \Big], \label{eq:sNihktolaphia} \\
   \hat{\sigma}_t(N_i h_k \rightarrow \ell_{\alpha} \Phi_a) &= \sum_{b,c=1}^{n_H} \frac{\mathbf{Y}_{\alpha i}^{b}\mathbf{Y}_{\alpha i}^{c\ast}\tilde{\mu}_{a b, k}\tilde{\mu}_{a c, k}^\ast}{32 \pi s} \Big\{\log \Big[m_{h_k}^2(m_{h_k}^2-2s) + (M_i^2-s)^2 \nonumber \\
    & + (s-M_i^2+m_{h_k}^2)\sqrt{\rho(s,M_i^2,m_{h_k}^2)} + 2 M_i^2 \Gamma_i^2\Big] \nonumber \\
    &- \log \Big[m_{h_k}^2(m_{h_k}^2-2s) + (M_i^2-s)^2  \nonumber \\
    &- (s-M_i^2+m_{h_k}^2)\sqrt{\rho(s,M_i^2,m_{h_k}^2)} + 2 M_i^2 \Gamma_i^2\Big]\Big\}, \label{eq:tRIS}
    \end{align}
    where $\rho(x,y,z)=(x-y-z)^2-4yz$ and the heavy neutrino total decay widths $\Gamma_i$ are shown in eq.~\eqref{eq:neutrinototal}. As mentioned in section~\ref{sec:BEs} and shown in diagram (\subref{fig:RIS2}) of figure~\ref{fig:diagsRIS}, for our 2RH neutrino case study of section~\ref{sec:model}, the $N_2$ mediated $s$-channel process $N_1 h_k \rightarrow \ell_{\alpha} \Phi_a$ contains a RIS that must be subtracted. Moreover, the Higgs mediated $t$-channel contribution also contains RIS. In fact, for this case, $N_i h_k \rightarrow \ell_{\alpha} \Phi_a$ can be decomposed as $N_i \rightarrow \ell_{\alpha} (\Phi_b \rightarrow \Phi_b) h_k \rightarrow \Phi_a$, where $\Phi_b$ is produced on-shell (see figure~\ref{fig:DeltaLeq1_scattering}). The first part corresponds to the heavy neutrino decay $N_i \rightarrow \ell_{\alpha} \Phi_b$ already accounted for in the BEs. To remove this RIS we follow the procedure outlined in ref.~\cite{Giudice:2003jh}. One needs to regulate the Higgs propagator via the external heavy neutrino $N_i$ decay width $\Gamma_i$, i.e. $t \rightarrow t + i M_i \Gamma_i$, which was noticed first in other contexts~\cite{Ginzburg:1995js,Melnikov:1996iu}. Upon integration, the result is a linearly divergent term $\propto 1/(M_i \Gamma_i)$, in the limit $M_i \Gamma_i \rightarrow 0$, corresponding to a Dirac delta function $\delta(t)$. This identifies an on-shell mediator corresponding to a RIS which is then removed. The above accounts for this subtraction. The $t$-channel is consistent with ref.~\cite{LeDall:2014too}. However, our $s$-channel result is distinct. Here, we obtain the correct "$+$" sign before the last term.

    The last of the $\Delta L =1$ diagrams corresponds to $N_i \Phi_a \to \ell_{\alpha} h_k$, labelled (\subref{fig:DeltaLeq1_Nphitohl}). We obtain
    \begin{align}
   \hat{\sigma}_{t_1}(N_i \Phi_a \to \ell_{\alpha} h_k)  &= \sum_{j,l=1}^{n_R}\dfrac{(1+\delta_{ij})(1+\delta_{il})|\mathbf{Y}_{\alpha i}|^2}{16\pi s(M_j^2-M_l^2)}\left\{\Big[(s-M_i^2)\mathbf{\Delta}_{ij}^{k*}\mathbf{\Delta}_{il}^{k}\right.\\
    &\left.-M_i(M_j\mathbf{\Delta}_{ij}^{k}\mathbf{\Delta}_{il}^{k}+M_l\mathbf{\Delta}_{ij}^{k*}\mathbf{\Delta}_{il}^{k*})\Big] \right.\nonumber\\
    &\left.\times\left(M_j^2\log\left[\dfrac{(s-M_i^2)(s-m_{h_k}^2)+sM_j^2}{sM_j^2}\right]-(j\leftrightarrow l)\right)\right.\nonumber\\
    &\left.-\mathbf{\Delta}_{ij}^{k}\mathbf{\Delta}_{il}^{k*}M_jM_l \right.\nonumber\\
    &\left.\times\left((s+M_j^2-m_{h_k}^2)\log\left[\dfrac{(s-M_i^2)(s-m_{h_k}^2)+sM_j^2}{sM_j^2}\right]-(j\leftrightarrow l)\right)\right\} , \nonumber\\
    \hat{\sigma}_{t_2}(N_i \Phi_a \to \ell_{\alpha} h_k) &= \sum_{b,c=1}^{n_H} \frac{\mathbf{Y}_{\alpha i}^{b}\mathbf{Y}_{\alpha i}^{c*}  \tilde{\mu}_{a b, k}\tilde{\mu}_{a c, k}^*}{32 \pi s}  \log \left[ \frac{s^2 (s-M_i^2-m_{h_k}^2)^2 + s^2 M_i^2 \Gamma_i^2}{ M_i^4 m_{h_k}^4 + s^2 M_i^2 \Gamma_i^2 } \right], 
    \end{align}
    where $t_1$ ($t_2$) refers to the $t$-channel diagram shown on the left (right) mediated by a heavy neutrino (Higgs). Once again the Higgs mediated $t$-channel contains a RIS, being the reason why the external heavy neutrino decay width $\Gamma_i$ appears in the expression above. We follow the same procedure described before for the $N_i h_k \to \ell_{\alpha} \Phi_a$ process to subtract this RIS [see eq.~\eqref{eq:tRIS}]. The quantity $\hat{\sigma}_{t_2}(N_i \Phi_a \to \ell_{\alpha} h_k)$ is in agreement with ref.~\cite{LeDall:2014too}. 
    
    In figure~\ref{fig:DeltaNeq2_scattering}, the diagrams (\subref{fig:DeltaNeq2_NNtohh}) correspond to the $N_i N_j \to h_p h_l$ process, we have
    \begin{align}
    \hat{\sigma}_s(N_i N_j \to h_p h_l) &=  \sum_{k,q=1}^{2 n_S} \frac{\tilde{\mu}_{kpl}\tilde{\mu}_{qpl}^*(1+\delta_{i j})^2}{16 \pi(s-m_{h_k}^2)(s-m_{h_q}^2)}  \frac{\sqrt{\rho(s,M_i^2,M_j^2)}}{s}\frac{\sqrt{\rho(s,m_{h_p}^2,m_{h_l}^2)}}{s} \nonumber \\ 
    &\times \left\{(s-M_i^2-M_j^2)\text{Re}\left[\mathbf{\Delta}^k_{i j}\mathbf{\Delta}^{q*}_{i j}\right] - 2 M_i M_j \text{Re}\left[\mathbf{\Delta}^k_{i j}\mathbf{\Delta}^q_{i j}\right] \right\} \nonumber \\ 
    & \times (1+\delta_{kp}+\delta_{kl}+\delta_{pl}+2\delta_{kp}\delta_{kl}\delta_{pl}) \nonumber \\ 
    & \times  (1+\delta_{qp}+\delta_{ql}+\delta_{pl}+2\delta_{qp}\delta_{ql}\delta_{pl})\;,
    \end{align}
    \begin{align}
    \hat{\sigma}_t(N_i N_j \to h_p h_l) &= \sum_{k,q=1}^{n_R} \frac{(1+\delta_{i k})(1+\delta_{jk})(1+\delta_{i q})(1+\delta_{j q})}{16 \pi s^2 (M_k^2-M_q^2)}\left\{-(M_k^2-M_q^2)\sqrt{\rho(s,M_i^2,M_j^2)}\right.\nonumber\\
    &\left.\times\sqrt{\rho(s,m_{h_p}^2,m_{h_l}^2)}\text{ Re}\left[\mathbf{\Delta}^{p}_{i k}\mathbf{\Delta}^{p*}_{i q}\mathbf{\Delta}^{l*}_{j k}\mathbf{\Delta}^{l}_{j q}\right]\right.\nonumber\\
    &\left.+\left[s\left(\log\left[(m_{h_l}^2-m_{h_p}^2)(M_i^2-M_j^2)\right.\right.\right.\right. \nonumber\\
    &\left.\left.+(m_{h_l}^2+m_{h_p}^2+M_i^2+M_j^2-2M_k^2-s)s-\sqrt{\rho(s,M_i^2,M_j^2)}\sqrt{\rho(s,m_{h_p}^2,m_{h_l}^2)}\right]\right.\nonumber\\
    &\left.\left. -\log\Big[(m_{h_l}^2-m_{h_p}^2)(M_i^2-M_j^2)+(m_{h_l}^2+m_{h_p}^2+M_i^2+M_j^2-2M_k^2-s)s \right.\right.\nonumber\\
    &\left.\left.+\sqrt{\rho(s,M_i^2,M_j^2)}\sqrt{\rho(s,m_{h_p}^2,m_{h_l}^2)}\Big]\right)\right.\nonumber\\
    &\left.\times\left(2M_iM_jM_k\left(M_q\text{ Re}\left[\mathbf{\Delta}^{p}_{i k}\mathbf{\Delta}^{p}_{i q}\mathbf{\Delta}^{l}_{j k}\mathbf{\Delta}^{l}_{j q}\right]+M_k\text{ Re}\left[\mathbf{\Delta}^{p*}_{i k}\mathbf{\Delta}^{p*}_{i q}\mathbf{\Delta}^{l}_{j k}\mathbf{\Delta}^{l}_{j q}\right]\right)\right.\right.\nonumber\\
    &\left.\left.+M_j(M_i^2+M_k^2-m_{h_p}^2)\left(M_k\text{ Re}\left[\mathbf{\Delta}^{p}_{i k}\mathbf{\Delta}^{p*}_{i q}\mathbf{\Delta}^{l}_{j k}\mathbf{\Delta}^{l}_{j q}\right]+M_q\text{ Re}\left[\mathbf{\Delta}^{p*}_{i k}\mathbf{\Delta}^{p}_{i q}\mathbf{\Delta}^{l}_{j k}\mathbf{\Delta}^{l}_{j q}\right]\right)\right.\right.\nonumber\\
    &\left.\left.+M_i(M_j^2+M_k^2-m_{h_l}^2)\left(M_k\text{ Re}\left[\mathbf{\Delta}^{p*}_{i k}\mathbf{\Delta}^{p*}_{i q}\mathbf{\Delta}^{l*}_{j k}\mathbf{\Delta}^{l}_{j q}\right]+M_q\text{ Re}\left[\mathbf{\Delta}^{p}_{i k}\mathbf{\Delta}^{p}_{i q}\mathbf{\Delta}^{l*}_{j k}\mathbf{\Delta}^{l}_{j q}\right]\right)\right.\right.\nonumber\\
    &\left.\left.+M_k M_q(M_i^2+M_j^2-s)\text{ Re}\left[\mathbf{\Delta}^{p*}_{i k}\mathbf{\Delta}^{p}_{i q}\mathbf{\Delta}^{l*}_{j k}\mathbf{\Delta}^{l}_{j q}\right]\right.\right.\nonumber\\
    &\left.\left.\left.+\left(M_i^2M_j^2+M_k^2(M_k^2-s)+m_{h_l}^2m_{h_p}^2-m_{h_l}^2(M_i^2+M_k^2)-m_{h_p}^2(M_j^2+M_k^2)\right) \right.\right.\right.\nonumber\\
    &\left.\left.\left.\times\text{Re}\left[\mathbf{\Delta}^{p}_{i k}\mathbf{\Delta}^{p*}_{i q}\mathbf{\Delta}^{l*}_{j k}\mathbf{\Delta}^{l}_{j q}\right]\right)-(k\leftrightarrow q)\right]\right\}\; , \\
    \hat{\sigma}_u(N_i N_j \to h_p h_p) &= \hat{\sigma}_t(N_i N_j \leftrightarrow h_p h_p)\; , \\
    \hat{\sigma}_{s-t}(N_i N_j \to h_p h_l) &=\sum_{k=1}^{n_R}\sum_{q=1}^{2n_S}\dfrac{(1+\delta_{ik})(1+\delta_{jk})(1+\delta_{ij})(1+\delta_{qp}+\delta_{ql}+\delta_{pl}+2\delta_{qp}\delta_{ql}\delta_{pl})}{8\pi s^2(s-m_{h_q}^2)}\nonumber\\
    &\times\text{ Re}[\tilde{\mu}_{qpl}]\left\{-\sqrt{\rho(s,m_{h_p}^2,m_{h_l}^2)}\sqrt{\rho(s,M_i^2,M_j^2)} 
    \right.\nonumber\\
    &\left.\times\left(M_i\text{ Re}\left[\mathbf{\Delta}^{q}_{i j}\mathbf{\Delta}^{p}_{i k}\mathbf{\Delta}^{l*}_{j k}\right]+M_j\text{ Re}\left[\mathbf{\Delta}^{q}_{i j}\mathbf{\Delta}^{p*}_{i k}\mathbf{\Delta}^{l}_{j k}\right]\right)\right.\nonumber\\
    &\left.+s\left(\log\left[(m_{h_l}^2-m_{h_p}^2)(M_i^2-M_j^2)+(m_{h_l}^2+m_{h_p}^2+M_i^2+M_j^2-2M_k^2-s)s\right.\right.\right.\nonumber\\
    &\left.\left.+\sqrt{\rho(s,M_i^2,M_j^2)}\sqrt{\rho(s,m_{h_p}^2,m_{h_l}^2)}\right]\right.\nonumber\\
    &\left.
    -\log\left[(m_{h_l}^2-m_{h_p}^2)(M_i^2-M_j^2)+(m_{h_l}^2+m_{h_p}^2+M_i^2+M_j^2-2M_k^2-s)s\right.\right.\nonumber\\
    &\left.\left.\left.-\sqrt{\rho(s,M_i^2,M_j^2)}\sqrt{\rho(s,m_{h_p}^2,m_{h_l}^2)}\right]\right)\right.\nonumber\\
    &\left.\times\left(2M_iM_jM_k\text{ Re}\left[\mathbf{\Delta}^{q}_{i j}\mathbf{\Delta}^{p}_{i k}\mathbf{\Delta}^{l}_{j k}\right]+M_k(M_i^2+M_j^2-s)\text{ Re}\left[\mathbf{\Delta}^{q*}_{i j}\mathbf{\Delta}^{p}_{i k}\mathbf{\Delta}^{l}_{j k}\right]\right.\right.\nonumber\\
    &\left.\left.+M_j(M_i^2+M_k^2-m_{h_p}^2)\text{ Re}\left[\mathbf{\Delta}^{q}_{i j}\mathbf{\Delta}^{p*}_{i k}\mathbf{\Delta}^{l}_{j k}\right]\right.\right.\nonumber\\
    &\left.\left.+M_i(M_j^2+M_k^2-m_{h_l}^2)\text{ Re}\left[\mathbf{\Delta}^{q*}_{i j}\mathbf{\Delta}^{p*}_{i k}\mathbf{\Delta}^{l}_{j k}\right]\right)\right\} \; , \\
    \hat{\sigma}_{s-u}(N_i N_j \to h_p h_p) &= \hat{\sigma}_{s-t}(N_i N_j \leftrightarrow h_p h_p)\; , 
    \end{align}
    \begin{align}
    \hat{\sigma}_{t-u}(N_i N_j \to h_p h_p) &= \sum_{k,q=1}^{n_R} \frac{(1+\delta_{i k})(1+\delta_{jk})(1+\delta_{i q})(1+\delta_{j q})}{8 \pi s^2 (2m_{h_p}^2+M_i^2+M_j^2-M_k^2-M_q^2-s)}\left\{-\text{Re}\left[\mathbf{\Delta}^{p}_{i k}\mathbf{\Delta}^{p*}_{i q}\mathbf{\Delta}^{p*}_{j k}\mathbf{\Delta}^{p}_{j q}\right]\right.\nonumber\\
    &\left.
    \times(2m_{h_p}^2+M_i^2+M_j^2-M_k^2-M_q^2-s)\sqrt{\rho(s,M_i^2,M_j^2)}\sqrt{\rho(s,m_{h_p}^2,m_{h_p}^2)}\right.\nonumber\\
    &\left.+\left[s\log\left[\dfrac{(2m_{h_p}^2+M_i^2+M_j^2-2M_k^2-s)s-\sqrt{\rho(s,M_i^2,M_j^2)}\sqrt{\rho(s,m_{h_p}^2,m_{h_p}^2)}}{(2m_{h_p}^2+M_i^2+M_j^2-2M_k^2-s)s+\sqrt{\rho(s,M_i^2,M_j^2)}\sqrt{\rho(s,m_{h_p}^2,m_{h_p}^2)}}\right]\right.\right.\nonumber\\
    &\left.\left.
    \times\Big(M_iM_j\left(2M_kM_q \text{Re}\left[\mathbf{\Delta}^{p}_{i k}\mathbf{\Delta}^{p}_{i q}\mathbf{\Delta}^{p}_{j k}\mathbf{\Delta}^{p}_{j q}\right]+(M_i^2+M_j^2-2m_{h_p}^2) \right.\right.\right.\nonumber\\
    &\left.\left.\left. \times\text{Re}\left[\mathbf{\Delta}^{p*}_{i k}\mathbf{\Delta}^{p*}_{i q}\mathbf{\Delta}^{p}_{j k}\mathbf{\Delta}^{p}_{j q}\right]\right) +M_j\left(M_k(m_{h_p}^2+2M_i^2+M_j^2-M_k^2-s) \right.\right.\right.\nonumber\\
    &\left.\left.\left. \times \text{Re}\left[\mathbf{\Delta}^{p}_{i k}\mathbf{\Delta}^{p*}_{i q}\mathbf{\Delta}^{p}_{j k}\mathbf{\Delta}^{p}_{j q}\right]+M_q(M_i^2+M_k^2-m_{h_p}^2) \text{Re}\left[\mathbf{\Delta}^{p*}_{i k}\mathbf{\Delta}^{p}_{i q}\mathbf{\Delta}^{p}_{j k}\mathbf{\Delta}^{p}_{j q}\right]\right)\right.\right.\nonumber\\
    &\left.\left.+M_i\left(M_k(m_{h_p}^2+M_i^2+2M_j^2-M_k^2-s)\text{ Re}\left[\mathbf{\Delta}^{p*}_{i k}\mathbf{\Delta}^{p*}_{i q}\mathbf{\Delta}^{p*}_{j k}\mathbf{\Delta}^{p}_{j q}\right]
    \right.\right.\right.\nonumber\\
    &\left.\left.\left.
    +M_q(M_j^2+M_k^2-m_{h_p}^2) \text{Re}\left[\mathbf{\Delta}^{p}_{i k}\mathbf{\Delta}^{p}_{i q}\mathbf{\Delta}^{p*}_{j k}\mathbf{\Delta}^{p}_{j q}\right]\right)\right.\right.\nonumber\\
    &\left.\left.
    +M_k M_q(M_i^2+M_j^2-s) \text{Re}\left[\mathbf{\Delta}^{p*}_{i k}\mathbf{\Delta}^{p}_{i q}\mathbf{\Delta}^{p*}_{j k}\mathbf{\Delta}^{p}_{j q}\right]\right.\right.\nonumber\\
    &\left.\left.
    +\left(M_i^2M_j^2-m_{h_p}^4-M_k^2(M_k^2-M_i^2-M_j^2-2m_{h_p}^2+s)\right)
    \right.\right.\nonumber\\
    &\left.\left. \times \text{Re}\left[\mathbf{\Delta}^{p}_{i k}\mathbf{\Delta}^{p*}_{i q}\mathbf{\Delta}^{p*}_{j k}\mathbf{\Delta}^{p}_{j q}\right]\Big)-(k\leftrightarrow q)\right]\right\}\; . 
    \end{align}
     The above scattering process was considered in ref.~\cite{AristizabalSierra:2014uzi} in the limit of a single heavy neutrino and one Yukawa coupling. Here we provide the complete expressions.
    
   Finally, diagrams (\subref{fig:DeltaNeq2_NNtohhphiphi}) contribute to $N_i N_j \to \Phi_a \Phi_b$, we have
    \begin{align}
    \hat{\sigma}_s(N_i N_j \to \Phi_a \Phi_b) &= \sum_{k,l=1}^{2 n_S} \frac{\tilde{\mu}_{a b, k}\tilde{\mu}_{a b, l}^*(1+\delta_{i j})^2}{16 \pi(s-m_{h_k}^2)(s-m_{h_l}^2)}  \frac{\sqrt{\rho(s,M_i^2,M_j^2)}}{s}  \nonumber \\ 
    & \times \left\{-2 M_i M_j \text{Re}\left[\mathbf{\Delta}^k_{i j}\mathbf{\Delta}^l_{i j}\right] +(s-M_i^2-M_j^2)\text{Re}\left[\mathbf{\Delta}^k_{i j}\mathbf{\Delta}^{l*}_{i j}\right] \right\}\; , \\
    \hat{\sigma}_t(N_i N_j \to \Phi_a \Phi_b) &=\sum_{\alpha,\beta=e,\mu,\tau}\dfrac{\mathbf{Y}^{a*}_{\alpha j}\mathbf{Y}^{a}_{\beta j}\mathbf{Y}^b_{\alpha j}\mathbf{Y}^{b*}_{\beta j}}{16\pi s}\Big\{\dfrac{s}{2}\log\left(\dfrac{s-M_i^2-M_j^2+\sqrt{\rho(s,M_i^2,M_j^2)}}{s-M_i^2-M_j^2-\sqrt{\rho(s,M_i^2,M_j^2)}}\right)\nonumber\\
    &-\sqrt{\rho(s,M_i^2,M_j^2)}\Big\} , \\
    \hat{\sigma}_{s-t}(N_i N_j \to \Phi_a \Phi_b) &=\sum_{\alpha=e,\mu,\tau}\sum_{k=1}^{2n_S} \dfrac{1}{16\pi s(s-m_{h_k}^2)}\left\{-\sqrt{\rho(s,M_i^2,M_j^2)} \right.\\
    &\left. \times \left(M_i\text{ Re}\left[\mathbf{Y}_{\alpha j}^b\mathbf{Y}_{\alpha i}^{a*}\tilde{\mu}_{a b, k}^*\mathbf{\Delta}^{k*}_{i j}\right]+M_j\text{ Re}\left[\mathbf{Y}_{\alpha j}^b\mathbf{Y}_{\alpha i}^{a*}\tilde{\mu}_{a b, k}^*\mathbf{\Delta}^{k}_{i j}\right]\right)\right.\nonumber\\
    &\left.+M_iM_j\log\left(\dfrac{s-M_i^2-M_j^2+\sqrt{\rho(s,M_i^2,M_j^2)}}{s-M_i^2-M_j^2-\sqrt{\rho(s,M_i^2,M_j^2)}}\right)\right.\nonumber\\
    &\left. \times\left(M_i\text{ Re}\left[\mathbf{Y}_{\alpha j}^b\mathbf{Y}_{\alpha i}^{a*}\tilde{\mu}_{a b, k}^*\mathbf{\Delta}^k_{i j}\right]+M_j\text{ Re}\left[\mathbf{Y}_{\alpha j}^b\mathbf{Y}_{\alpha i}^{a*}\tilde{\mu}_{a b, k}^*\mathbf{\Delta}^{k*}_{i j}\right]\right)\right\}. \nonumber
    \end{align}
    The $s$-channel contribution was already computed in ref.~\cite{LeDall:2014too} and our results are consistent. Furthermore, the $t$-channel occurs in vanilla type-I seesaw leptogenesis and was already computed in another context in ref.~\cite{Plumacher:1998ex} and our results are also in agreement.

\end{document}